\documentclass[12pt,preprint]{aastex}

\newcommand{\Lya}{\mbox{\rm Ly-$\alpha$}}
\newcommand{\Lyb}{\mbox{\rm Ly-$\beta$}}
\newcommand{\kms}{\mbox{\rm km\thinspace s$^{-1}$}}
\newcommand{\Anaught}{\mbox{${\cal A}_0$}}
\newcommand{\FWHM}{\mbox{\rm FWHM}}
\newcommand{\Wstar}{\mbox{$W_\star$}}
\newcommand{\Wrest}{\mbox{$W_{\lambda_0}$}}
\newcommand{\zabs}{\mbox{$z_{\rm abs}$}}
\newcommand{\zem}{\mbox{$z_{\rm em}$}}
\newcommand{\Vej}{\mbox{$V_{\rm ej}$}}
\newcommand{\NHI}{\mbox{$N_{\mbox{\scriptsize\rm H\thinspace{\sc i}}}$}}

\begin{document}


\title{A uniform analysis of the Ly-$\alpha$ forest at $z = 0-5$:\\
IV. The clustering and evolution of clouds at $z \leq 1.7$.}

\author{Adam Dobrzycki}
\affil{Harvard-Smithsonian Center for Astrophysics, 60 Garden Street,
MS 70 \\ Cambridge, MA 02138, USA \\
e-mail: adobrzycki@cfa.harvard.edu}

\author{Jill Bechtold, Jennifer Scott, Miwa Morita}
\affil{Steward Observatory, University of Arizona \\ Tucson, AZ 85721,
USA \\
e-mail: [jbechtold,jscott,mmorita]@as.arizona.edu}

\shorttitle{\Lya\ clouds at $z \leq 1.7$}
\shortauthors{Dobrzycki et al.}

\slugcomment{Accepted for publication in Ap.J.}


\begin{abstract}

We present results on the evolution and clustering of \Lya\ lines at
low redshift as part of our series ``A uniform analysis of the \Lya\
forest at $z = 0-5$.''  The sample analyzed in this paper contains
1298 \Lya\ absorption lines from 165 quasar spectra mined from the
archives of the Faint Object Spectrograph on the Hubble Space
Telescope (HST). Our sample extends to $z\approx1.7$, slightly higher
than the sample analyzed by the HST Quasar Absorption Line Key
Project. We confirm the result from the Key Project that the number
density evolution of \Lya\ lines at low redshifts can be described by
a power law that is significantly flatter than that found at high
$z$. We find that the evolution is somewhat steeper than obtained
previously. Specifically, we find $\gamma = 0.54 \pm 0.21$ for lines
with equivalent widths greater than 0.24~\AA\, and $\gamma=0.60 \pm
0.14$ using a variable equivalent width threshold. We find that the
difference between our and Key Project results is likely attributable
to different redshift coverage of the two samples. The results
concerning the number density evolution are not significantly affected
if one includes \Lya\ lines which are members of metal systems. Object
to object fluctuations in the number of lines detected are small,
indicating a high degree of uniformity in the intergalactic medium on
large scales. We find marginal evidence that weak and strong lines
undergo different evolution. We find weak clustering for \Lya\ lines
at velocity separations $\Delta V \leq\ 500$~\kms, weaker than the
level predicted from an earlier analysis by Ulmer of a small subsample
of the Key Project data. We see no correlations for metal
system--\Lya\ forest or extensive metal system--\Lya\ forest
combinations.

\end{abstract}




\section{Introduction}

This paper is a continuation of the project of the analysis of \Lya\
absorption lines covering the range of redshifts, $z=0-5$. The first
two papers in this series (Scott, Bechtold, \& Dobrzycki 2000; Scott
et al.\ 2000; hereafter Papers~I and II) presented the analysis of
intermediate resolution ground based data. This and accompanying
papers (Bechtold et al.\ 2001, Paper~III; Scott et al.\ 2001, Paper~V)
report the analysis of Hubble Space Telescope (HST) Faint Object
Spectrograph (FOS; Ford \& Hartig 1990) archival spectra.

Paper~III describes the FOS data reductions, absorption line analysis,
and line identification procedure. In this paper we present the
distribution and clustering properties of the \Lya\ absorption lines.

All data analyzed in this and other papers from this series can be
accessed from our web pages:\\
\verb@      http://lithops.as.arizona.edu/~jill/QuasarSpectra/@\\
or\\
\verb@      http://hea-www.harvard.edu/QEDT/QuasarSpectra/@.


\section{The Lyman $\alpha$ absorption line sample}
\label{sec-sample}

From data presented in Paper~III, we selected all spectra suitable for
the analysis of the \Lya\ forest. We excluded spectra observed with
A-1 aperture before installation of COSTAR, broad-absorption line
quasars, and quasars observed in spectropolarimetry mode (see
Paper~III). In cases where two images of a gravitationally lensed
quasar were observed, or a close pair of quasars or grouping of
quasars were observed, we included only one line-of-sight, choosing
the spectra with the highest signal-to-noise.

For detailed discussion on the detection significance and completeness
in our sample, and on the identification of metal lines, see Sec.~2 of
Paper~III.

In total, the sample analyzed in this paper contains spectra of 165
quasars (Table~\ref{tab:quasars}). The spectral resolution of all the
data is 230--280~\kms\ \FWHM. Of those 165 quasars, 63 were analyzed
by Weymann et al.\ (1988) as part of the HST Quasar Absorption Line
Key Project (Bahcall et al.\ 1993, 1996; Schneider et al.\ 1993,
Jannuzi et al.\ 1998, Weymann et al.\ 1998). The sixty fourth object
in Weymann et al.\ sample, PG~2302+029, is a broad absorption line
quasar (Jannuzi et al.\ 1996) and was excluded from our sample (as
well as from Samples 6--10 in Weymann et al.).

We selected the sample of the absorption lines for further analysis
using the following criteria:

1. The detection significance for the line has to be greater than
5$\sigma$. 

2. A line cannot be identified as a heavy element absorption
feature. Regions within 300~\kms\ of the centers of metal absorption
lines were excluded from the analysis.

3. Whenever a \Lya\ ``forest'' sample was called for, we excluded
\Lya\ lines identified as members of heavy element systems and regions
within 300~\kms\ from such lines.

4. We excluded lines lying bluewards from the location of the \Lyb\
emission line, to avoid confusion with the \Lyb\ lines from higher
redshift \Lya\ systems.

\begin{figure}
\plotone{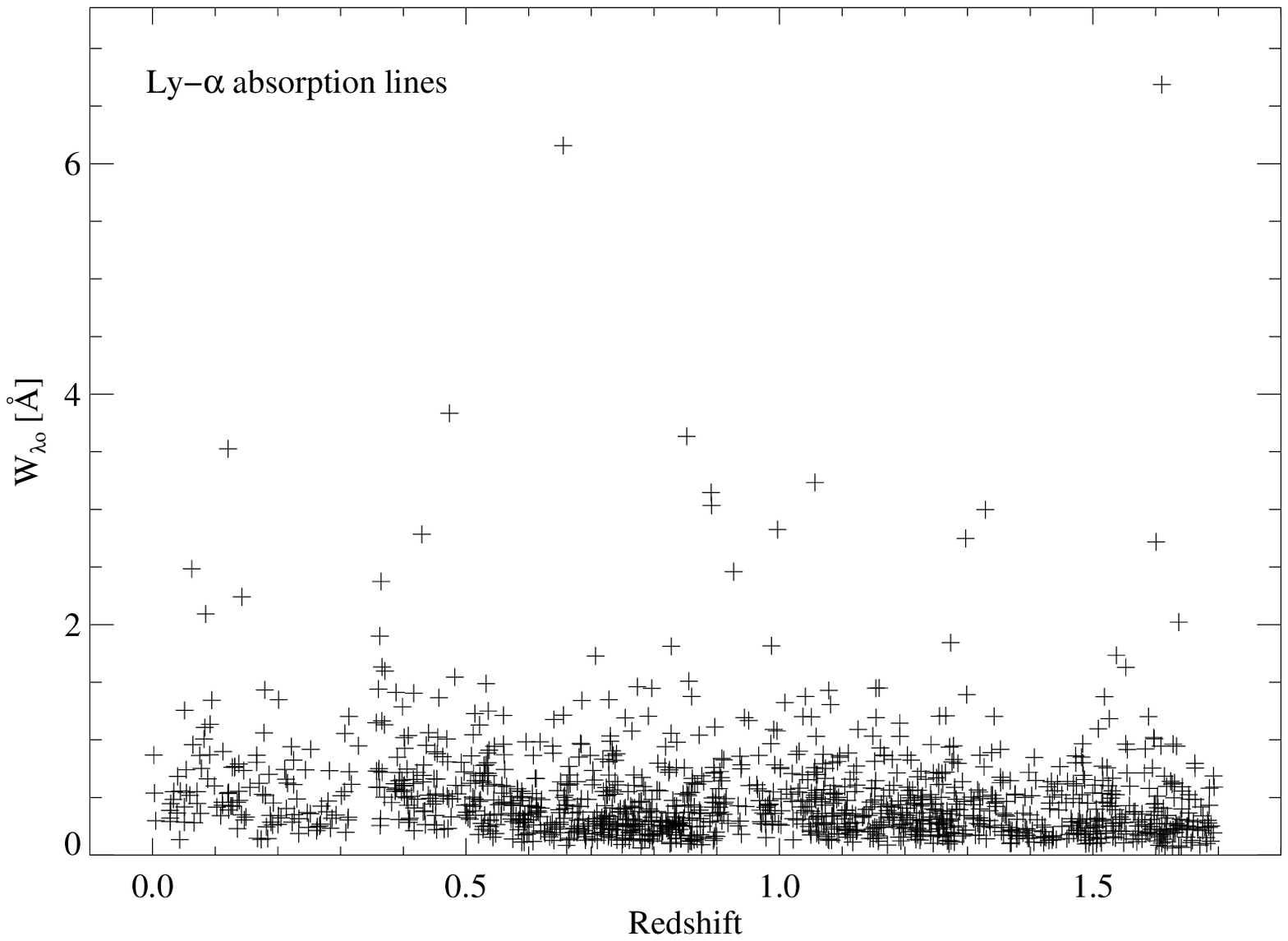}
\caption{Equivalent widths of all \Lya\ lines, forest and metal
systems, versus redshift.\label{fig:eqw}}
\end{figure}

Please note that these criteria are more conservative than those of
the HST Key Project. The sample contains 1298 lines if all \Lya\ lines
(forest and metal systems) are selected, and 1157 \Lya\ forest lines.
Figure~\ref{fig:eqw} shows the rest equivalent width, \Wrest, defined
as $\Wrest = W_{\lambda{\rm (obs)}}/(1+z)$, of all \Lya\ lines versus
redshift, $z$. Equivalent widths were measured from gaussian fits to
significant absorption features (see Paper~III). This plot has not
been corrected for the relative sensitivity for detection at each
redshift and is only intended to show the redshift distribution of the
sample.

\begin{figure}
\plotone{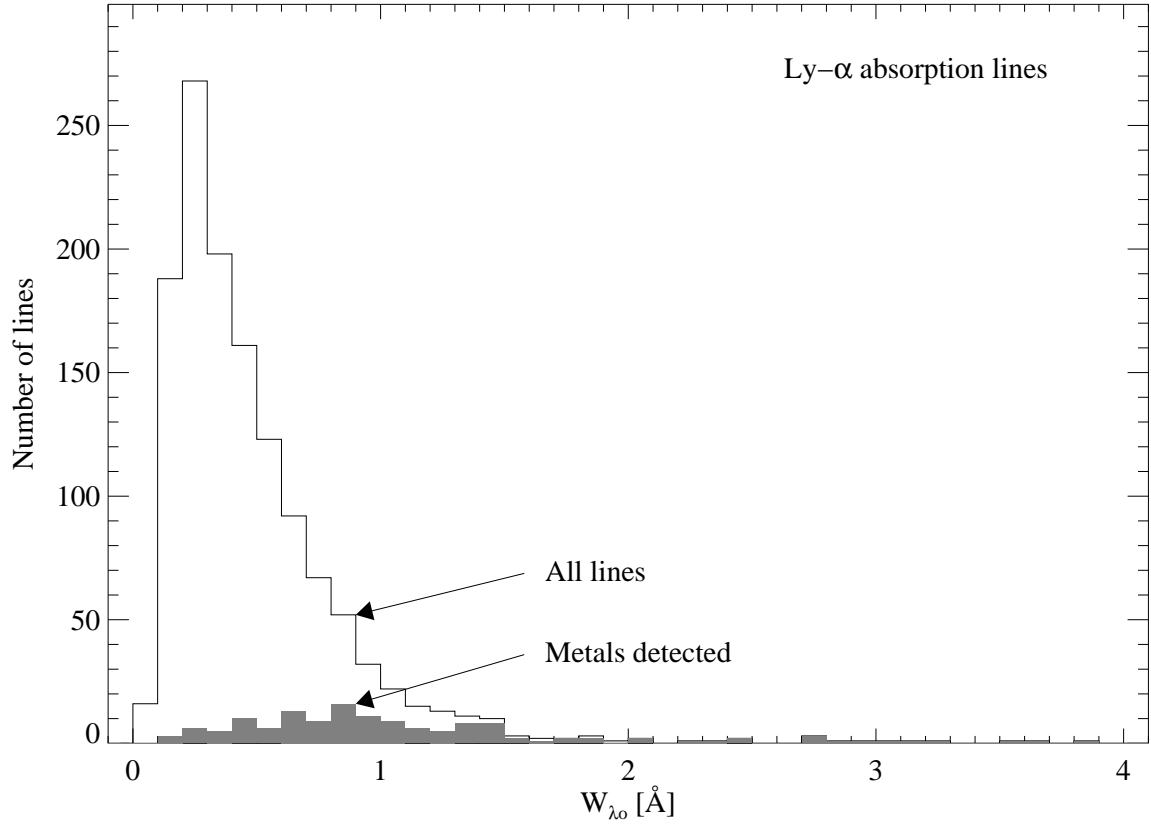}
\caption{The histogram of rest equivalent widths for \Lya\ lines.
The shaded area shows the fraction of \Lya\ lines from heavy element
systems.\label{fig:eqwhist}}
\end{figure}

Figure~\ref{fig:eqwhist} shows the histogram of rest equivalent widths
of lines in the full \Lya\ sample. The shaded area shows the \Lya\
lines with detected metal absorption. Figure~\ref{fig:eqwhist} clearly
shows that lines associated with metal systems begin to compose a
significant part of the sample for rest equivalent widths greater than
$\sim$1~\AA. We cannot rule out the possibility that some of the lines
in our sample are in fact unidentified metal system lines. However, we
found that the exclusion from the analysis of all lines with $\Wrest >
1$~\AA\ --- which, in view of Fig.~\ref{fig:eqwhist}, are the prime
suspects for being unrecognized metal lines --- produced negligible
differences in the number density evolution results.

\begin{figure}
\plotone{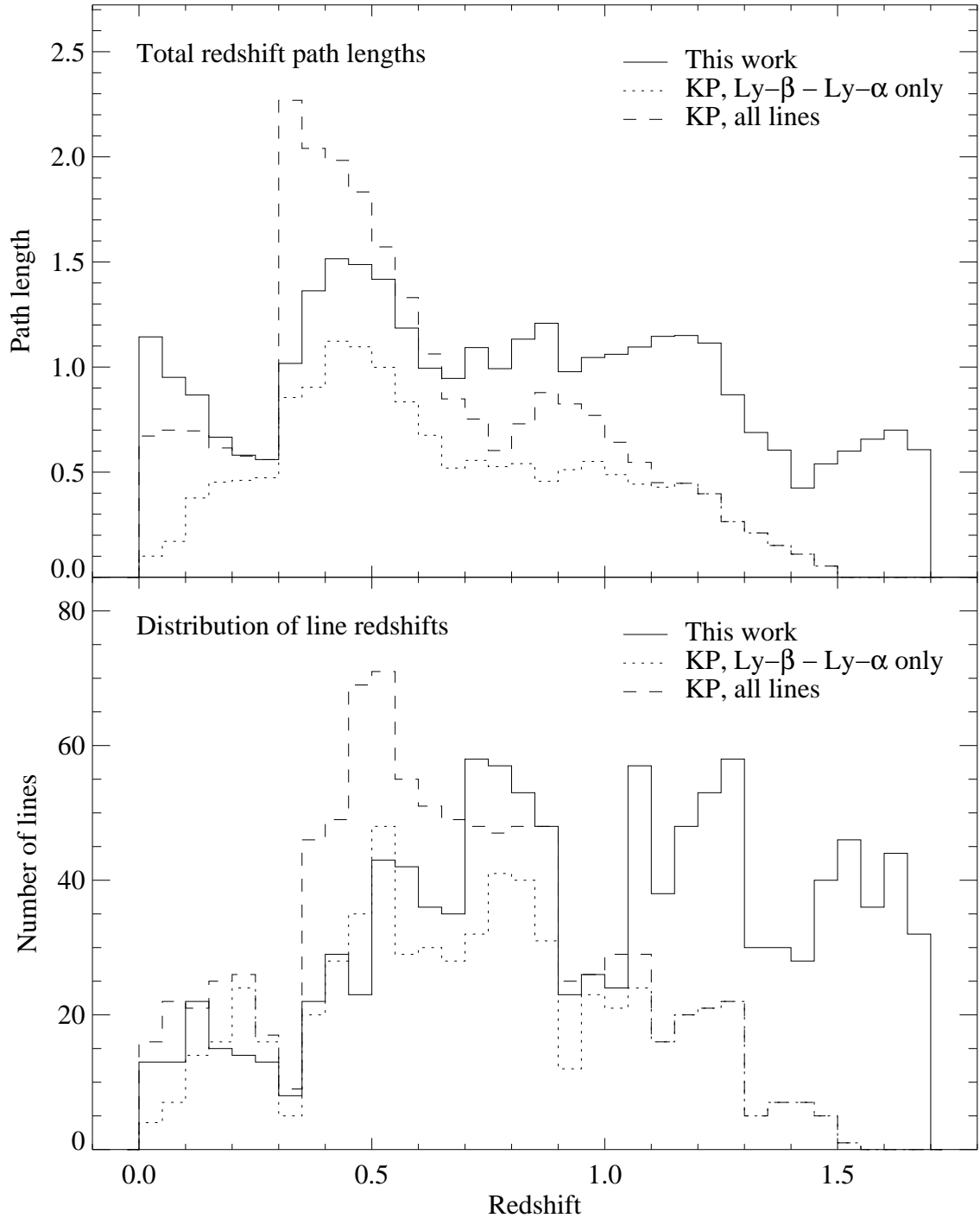}
\caption{Histogram of total redshift path length (upper panel) and
lines (lower panel) versus redshift in our full \Lya\ forest sample
(solid line), Weymann et al.\ (1988) full sample (dashed line) and
Weymann et al.\ \Lyb\--\Lya\ only sample (dotted
line).\label{fig:zcoverage}}
\end{figure}

Figure~\ref{fig:zcoverage} shows the comparison of redshift coverage
in our sample with the Weymann et al.\ (1998) sample. The upper panel
shows the combined path length available for study in both samples as
a function of redshift, while the lower panel shows the number of
available lines. The notable differences are that (1) our sample
extends to $z \sim 1.8$, while the Key Project sample effectively ends
at $z \sim 1.5$, (2) our sample contains considerably more lines in
the $z=1.0-1.5$ range, and (3) our sample by definition excludes lines
below \Lyb\ emission line, while the Key Project sample does not.

We note that did not attempt to remove the flat field features in the
FOS spectra, which the Key Project went to great lengths to do
(Schneider et al.\ 1993, Jannuzi et al.\ 1998). We identify 38 of
their 141 confirmed flat field features as lines with significance
greater than 5$\sigma$ (Paper~III). However, only nine of those
features lie in the \Lya\ forest. Simple extrapolation of those
numbers to the sample analyzed here would indicate that roughly 24
lines out of our total sample of 1298, or less than 2\%, are
spurious. However, even this low number is likely overestimated, since
majority of redshift path in our sample that is added to the Key
Project sample (see Fig.~\ref{fig:zcoverage}) is in regions covered by
G130H and G270H gratings, which have less (none in the case of G130H,
actually) identified flat field features than grating G190H (Jannuzi
et al.\ 1998). Those should have a negligible effect on our results.


\section{Properties of Ly-$\alpha$ absorbers at $z\lesssim1.7$}
\label{sec-dndz}

We parameterized the distribution of individual clouds using:
\begin{equation}
\frac{\partial^2{\cal N}}{\partial z\,\partial\Wrest} =
\Anaught\Wstar^{-1}
\left(1+z\right)^\gamma \exp\left(-\frac{\Wrest}{\Wstar}\right)
\label{eq:dndz}
\end{equation}
where \Anaught, $\gamma$, and \Wstar\ are the distribution
parameters. For a non-evolving population of clouds $1/2 \leq \gamma
\leq 1$, depending on the value of $q_0$ (Peterson 1978). It is
well-established that $\gamma$ is significantly greater than 1 at $z
\gtrsim 1.7$, indicating strong evolution (e.g.\ Bajtlik, Duncan, \&
Ostriker 1988; Lu, Wolfe, \& Turnshek 1991; Bechtold 1994; Kim et al.\
1997; Paper~I). However, the HST Quasar Absorption Line Key Project
and other studies showed that $\gamma$ is less then 1 at $z \lesssim
1.7$, consistent with the no evolution case (Bahcall et al.\ 1993,
1996; Weymann et al.\ 1998; Vanden Berk et al.\ 1999; Impey, Petry, \&
Flint 1999).

The proximity effect (Bajtlik, Duncan \& Ostriker 1988; Bechtold 1994;
Paper~II) at high redshift causes the derived value of $\gamma$ to
depend strongly on the inclusion or exclusion of lines with $\zabs
\sim \zem$. However, the proximity effect is not expected to influence
the results as significantly at low redshift for a few reasons. First,
the overall number density of absorbers per unit redshift is smaller
than at high redshift, and so the expected number of lines affected
will also be small. Second, the quasars used at low redshift are
intrinsically fainter on average than the quasars observed at high
$z$, and thus it is expected that their influence at their environment
will also be less prominent (see Paper V). We determined the
distribution parameters both including and excluding lines with
``ejection velocity'' \Vej, velocity displacement from the emission
redshift of the quasar, greater than 3000~\kms. A detailed analysis of
the proximity effect is presented in Paper~V.

\subsection{The method}

We estimated the parameters \Anaught, $\gamma$, and \Wstar\ from
Eq.~(\ref{eq:dndz}) for the whole FOS sample and several subsamples.
We developed and utilized a computer code, based on the algorithm from
Murdoch et al.\ (1986; their equations [A2a] and [A8a] corrected),
modified so that the integral in Eq.~(A2) of Murdoch et al., which
cannot be calculated analytically in the case of variable threshold,
is derived numerically. The Murdoch et al.\ method consists of solving
for $\gamma$ and \Wstar\ by finding the roots of the derivatives of
the log of the likelihood function describing Equation~(\ref{eq:dndz})
with respect to $\gamma$ and \Wstar. Uncertainties in each parameter
are found from a parabolic fit to $\ln L$ such that 1$\sigma$
corresponds to $\ln L =\ln L_{\rm max}-1/2$. The normalization,
\Anaught, is calculated by dividing the total number of lines by the
sum of the integral of Equation~(\ref{eq:dndz}) over the redshift
paths in all quasars:
\begin{equation}
\Anaught={\cal N} \times 
\left[\sum_{q} \int\limits_{z^{q}_{\rm min}}^{z^{q}_{\rm max}} 
\left(1+z\right)^\gamma dz\right]^{-1}.
\label{eq:anaughtintegral}
\end{equation}
for the considered sample. For each run of our program, we tested the
goodness-of-fit of the outcome using Kolmogorov-Smirnoff test.

This treatment of the normalization differs slightly from that of the
HST Key Project. We and the Key Project both solve for $\gamma$ and
\Wstar\ using all lines above some rest equivalent width that is
either fixed to some constant value or varied according to the
signal-to-noise ratio in each spectral region. Weymann et al.\ (1998)
determine the normalization, denoted $dN/dz$ in their Table~2,
explicitly in the maximum likelihood solution rather than by using
Eq.~(\ref{eq:anaughtintegral}). Each $dN/dz$ value listed in their
table is the local line density relative to the local density of lines
above 0.24~\AA, regardless of the limiting equivalent width of the
sample used.  In our software, this relative scaling is not done. In
other words, we always normalize to the stated constant rest
equivalent width limit and do not quote a value of \Anaught\ for the
case of a variable rest equivalent width limit.

\subsection{Results}

\subsubsection{Number density evolution}

Table~\ref{tab:dndz-hst} presents a summary of our estimates of the
parameters from Eq.~(\ref{eq:dndz}) in several subsamples of the data
set. The main result, seen clearly in Table~\ref{tab:dndz-hst}, is
that in all cases the value of the evolution index, $\gamma$, is
within few tenths of 0.5, significantly lower than the value estimated
from ground-based samples of \Lya\ forest lines at redshifts greater
than $\sim$1.7. Our values are consistent with the ``no evolution''
expectations. We show the results for the $\Wrest \geq 0.24$~\AA,
$\Vej \geq 3000$~\kms\ sample in Figure~\ref{fig:dndzuskp}. The values
of the K-S probability show that the power law is an excellent
approximation of the distribution of lines at the redshift range
covered in our sample. Note that this disagrees with the result of
Kim, Cristiani, \& D'Odorico (2001), who found that the break in
$d{\cal N}/dz$ slope occurred around $z \sim 1.2$.

The results appear to depend only weakly on whether the \Lya\ lines
from metal systems are included in the sample. Also, it can be seen in
Table~\ref{tab:dndz-hst} that taking the proximity effect into
account --- by excluding the lines with $\Vej < 3000$~\kms\ ---
results in an insignificant change in the estimates of the number
density evolution index; the difference is of the order of 1$\sigma$
or less.

\subsubsection{Comparison with HST QSOAL Key Project}

The results agree qualitatively with the findings of the HST QSOAL Key
Project. Weymann et al.\ (1998) have analyzed the complete Key Project
\Lya\ absorption line sample and also found that the evolution index,
$\gamma$, is smaller than the one found for high-redshift
samples. However, their values of $\gamma$ for various \Lya\
subsamples range from $-0.03$ to $+0.26$ with uncertainties of
$\sim0.20$. The values of $\gamma$ that we find from our total FOS
sample are larger than those found by the Key Project by
$\sim$1.9$\sigma$ in the case of the constant, 0.24~\AA\ equivalent
width threshold and by $\sim$4.5$\sigma$ in the case of a variable
equivalent width threshold

\begin{figure}
\plotone{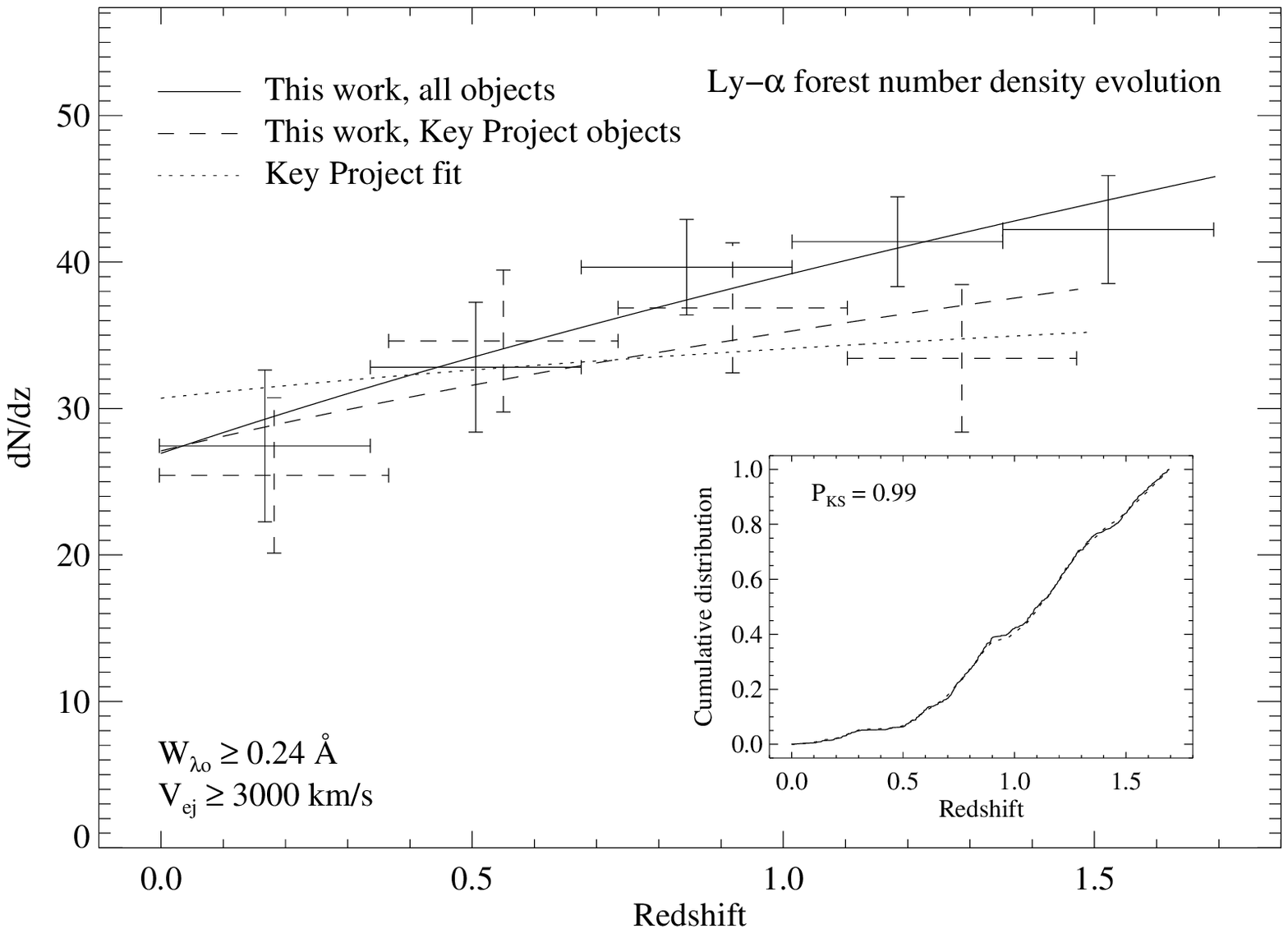}
\caption{Number density evolution of \Lya\ lines for the
$\Wrest\geq0.24$~\AA, $\Vej\geq3000$~\kms\ sample. Solid data points
and line: our results for the whole sample. Dashed data points and
curve: our results for the sample limited to the Weymann et al.\
(1988) objects. Dotted line: fit from Weymann et al.\ (1998), their
sample \#5. The data are binned for presentation only; the lines show
the fits to unbinned data. The inner box shows the comparison of
cumulative distribution of lines in the sample (solid line) and
Eq.~(\ref{eq:dndz}) integrated over the sample (dashed
line).\label{fig:dndzuskp}}
\end{figure}

A comparison of these two results is shown in
Figure~\ref{fig:dndzuskp}, where we show our data and fit for
$\Wrest\geq0.24$~\AA\ (solid line), and the $d{\cal N}/dz$ vs.\ $z$
curve (dotted line) for a comparable sample from Weymann et al.\
(1998; their sample \#5, which had $\Anaught=30.7$ and $\gamma = 0.15
\pm 0.23$). We see a comparable number of lines per unit redshift at
low $z$, but slightly (ca.\ 15\%) more lines at $z \approx 1.5$.

There are a few effects that may account for the discrepancy. As noted
earlier (see Fig.~\ref{fig:zcoverage}), our sample has larger redshift
coverage and contains markedly more lines at $z=1.0-1.5$. Also, our
sample excludes lines below \Lyb\ emission line.

We determined the parameters of Eq.~(\ref{eq:dndz}) in a subsample of
our data, containing only the objects we had in common with the Key
Project.  The results were in very good agreement with Weymann et al.\
(1998); we show the data points and the fit on
Fig.~\ref{fig:dndzuskp}. Specifically, we obtained $\gamma = 0.38 \pm
0.36$ for the $\Wrest \geq 0.24$~\AA\ sample, and $\gamma = 0.26 \pm
0.22$ for the variable threshold case, markedly lowering any
discrepancy. Also, the results show the same qualitative behavior of
lowering the value of $\gamma$ when limiting the analysis to the Key
Project objects.

The other effect that can influence the results is the exclusion of
lines below \Lyb\ emission lines. In order to verify whether this can
account for at least part of the discrepancy, we analyzed the \Lya\
line lists of the HST Key Project using our equivalent width threshold
information and our metal system identifications for excluding
spectral regions. Including identified \Lya\ lines blueward of \Lyb\
emission --- as the Key Project did --- we find $\gamma = -0.03 \pm
0.20$ and $\gamma=-0.10 \pm 0.29$, for variable and 0.24~\AA\
thresholds, respectively, in good agreement with the Key Project
results. Excluding lines blueward of \Lyb\ emission, we find $\gamma =
-0.03 \pm 0.24$ and $\gamma = -0.01 \pm 0.32$ for variable and
0.24~\AA\ thresholds, respectively, i.e.\ no significant difference in
both cases. We thus conclude that exclusion or inclusion of lines
below \Lyb\ does not affect the results, and that the primary source
of the discrepancy between our and Weyman et al.'s (1998) results lies
in the difference in the sample coverage.

It is known that $d{\cal N}/dz$ steepens around $z\sim1.6-1.8$. It is
obvious from Fig.~5 of Weymann et al.\ (1998) that if they included
the $\zem = 1.9$ quasar UM18 in their sample (it was excluded because
of incomplete line identifications; note that this object is included
in the sample presented here), then their derived distribution would
have been considerably steeper, in agreement with our result. It is
possible that this, combined with the fact that we have better
coverage at high-$z$ end of the sample, contributes to our deriving
larger $\gamma$ than the Key Project. We note, however, that the fact
that K-S test shows that for our sample the power law is an excellent
approximation of the number density evolution over the entire redshift
range of our sample is in apparent contradiction to what inclusion of
UM18 would do to the Key Project sample, in which case their fit to a
single power law would likely become worse.

\subsubsection{Dependence of evolution on line equivalent width}

If the intergalactic medium evolves, then we expect that the observed
$\gamma$ could depend on equivalent width, or equivalently, the
distribution of equivalent widths could be a function of redshift.
Weymann et al.\ (1998) found a trend of increasing $\gamma$ with
increasing line strength in the Key Project data. However, Penton,
Shull, \& Stocke (2000) saw no difference in the evolution of strong
and weak lines in the HST/GHRS sample.

\begin{figure}
\plotone{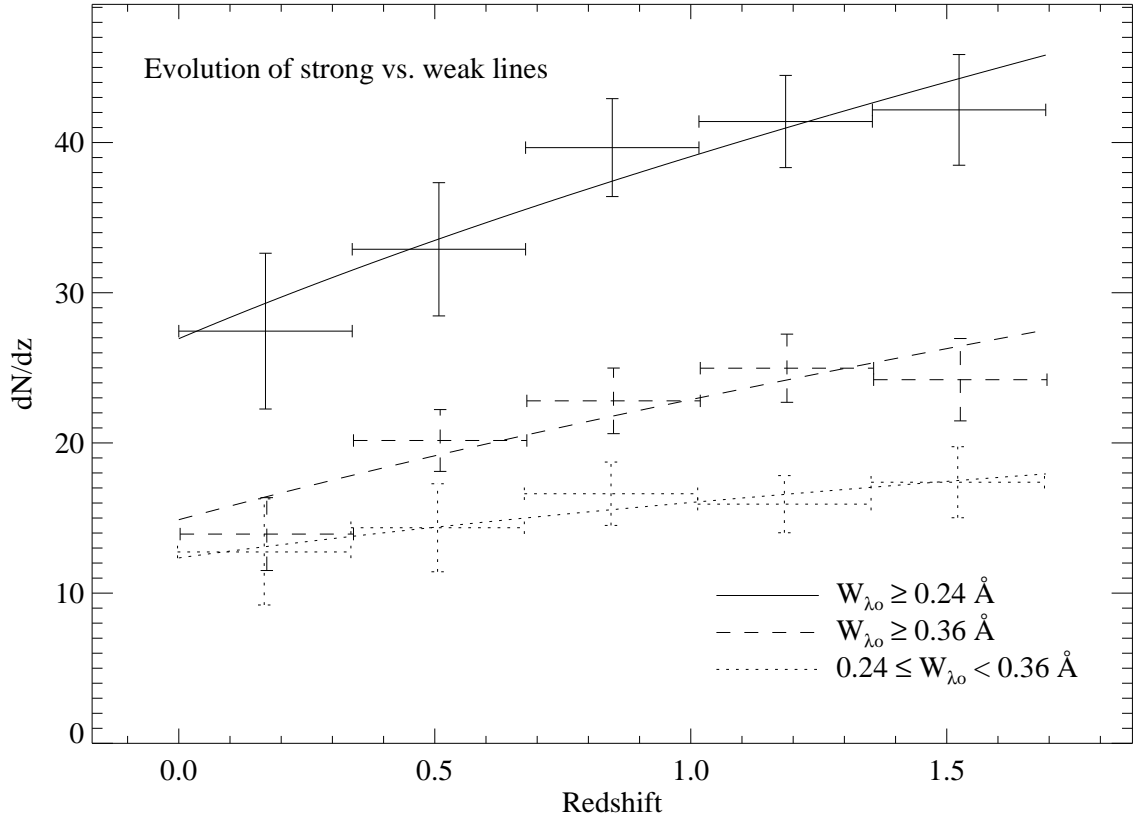}
\caption{Number density evolution as a function of equivalent
width. Solid lines: lines with $\Wrest \geq 0.24$~\AA, dashed lines:
absorbers with $\Wrest \geq 0.36$~\AA, dotted lines: absorbers with
$0.24 \leq \Wrest < 0.36$~\AA. Note marginally different slopes for
weak and strong lines. Data are binned for presentation; the fits are
to unbinned data.\label{fig:hstdndz}}
\end{figure}

Although the statistical significance is low, our results suggest that
there appears to be a difference in the rate of evolution for weak and
strong lines. In Figure~\ref{fig:hstdndz} the solid line shows the
evolution of the number density of absorbers for lines from one of the
samples, with $\Wrest \geq 0.24$~\AA. On the same plot, we show the
data and fits for two of its subsamples: the dashed line shows the
sample with $\Wrest \geq 0.36$~\AA, the dotted line --- $0.24 \leq
\Wrest <0.36$~\AA. As in other plots, the data have been arbitrarily
binned solely for presentation; the lines show the fits to unbinned
data.

While number densities of strong and weak lines seem to be comparable
in the present epoch, there are more (by roughly 25\%) strong lines at
$z\approx1.7$. In principle, the dividing line between those two
subsamples lies near the point where \Lya\ lines leave the linear part
of the curve of growth. Therefore, one could try to interpret this
result as an indication that the two subsamples are in fact two
different populations of absorbers. However, there are two reasons why
we regard this result with caution and avoid drawing far-reaching
conclusions. First, the result is marginal. Second, the same behavior
would be expected to occur from line blending. Two blending effects
associated with increased density of lines at larger $z$ may affect
the relative number of strong and weak lines: (1) two lines, both
above the threshold, may blend into features indistinguishable from
strong lines (``blending-out''), or (2) two weak lines, both with
\Wrest\ below the threshold, may blend, resulting in a feature that
has combined width greater than the threshold (``blending-in''). These
two effects work in opposite directions, and they are hard to
quantify, especially in spectra with limited resolution (see, e.g.,
Parnell \& Carswell 1988, Liu \& Jones 1988). If the first of those
two effects dominates over the second, the result may appear as if
there were more strong lines at higher redshifts.

\subsubsection{Variation in $d{\cal N}/dz$ from quasar to quasar}

\begin{figure}
\plotone{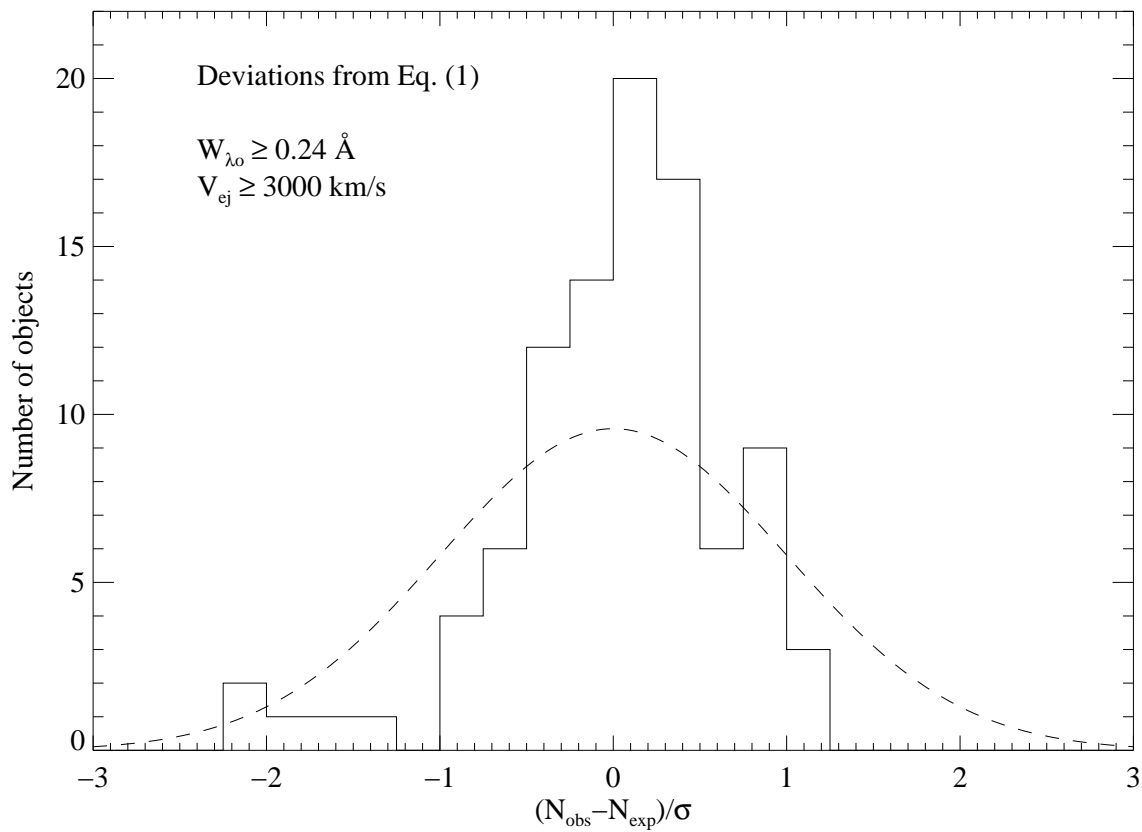}
\caption{The histogram of deviations of observed number of lines
in individual objects from the expected number of lines, expressed in
units of individual uncertainties for each object. The expected number
of lines was calculated by integrating Eq.~(\ref{eq:dndz}) over the
available regions of spectra. Dashed line shows a gaussian
distribution with unit $\sigma$, normalized to the size of the
sample.\label{fig:dndzfluct}}
\end{figure}

Impey, Petry, \& Flint (1999) reported marginal evidence for
quasar-to-quasar variations in the number density of lines in their
sample of 10 spectra taken with HST/GHRS. Our sample, which contains
165 lines of sight, is particularly well suited to test whether there
is structure in the distribution of \Lya\ absorbers on very large
scales. We find remarkably little scatter in the number of lines
detected. For each object, we calculated the expected number of
absorption lines by integrating Eq.~(\ref{eq:dndz}) over the usable
parts of the spectrum, and compared it with the number of lines
observed in the spectrum.  Figure~\ref{fig:dndzfluct} shows the
histogram of the deviations in units of the individual
uncertainties. Since the number of lines in each object can be small,
the individual uncertainties were calculated using the Gehrels (1986)
approximation, $\sigma_{\cal N} = 1 + \sqrt{{\cal N}+3/4}$, where
${\cal N}$ is the predicted number of lines. If the distribution were
random, then the histogram would be a gaussian with unit sigma. A
dotted line in Figure~\ref{fig:dndzfluct} shows the predicted gaussian
distribution. The observed distribution is in fact much {\em narrower}
than random: the farthest outliers deviate by only $2\sigma$, and
there are only a handful of objects outside $\pm 1 \sigma$.

The explanation is that the deviations in the number of lines detected
in the FOS spectra is not given by counting statistics. Rather, the
deviations reflect the uniformity of the intergalactic medium averaged
over large scales. The size scale probed is given roughly by the
$\Delta z$ between \Lya\ and \Lyb\ emission lines of each quasar, or
$\sim$750~$h_{50}^{-1}$~Mpc at $z=1$ and $\sim$850~$h_{50}^{-1}$~Mpc
at $z=0.5$ ($q_0=0.5$).

We will explore the uniformity of the line distribution in more detail
in a future paper.

\subsubsection{Line equivalent width and neutral hydrogen column
density distributions}

\begin{figure}
\plotone{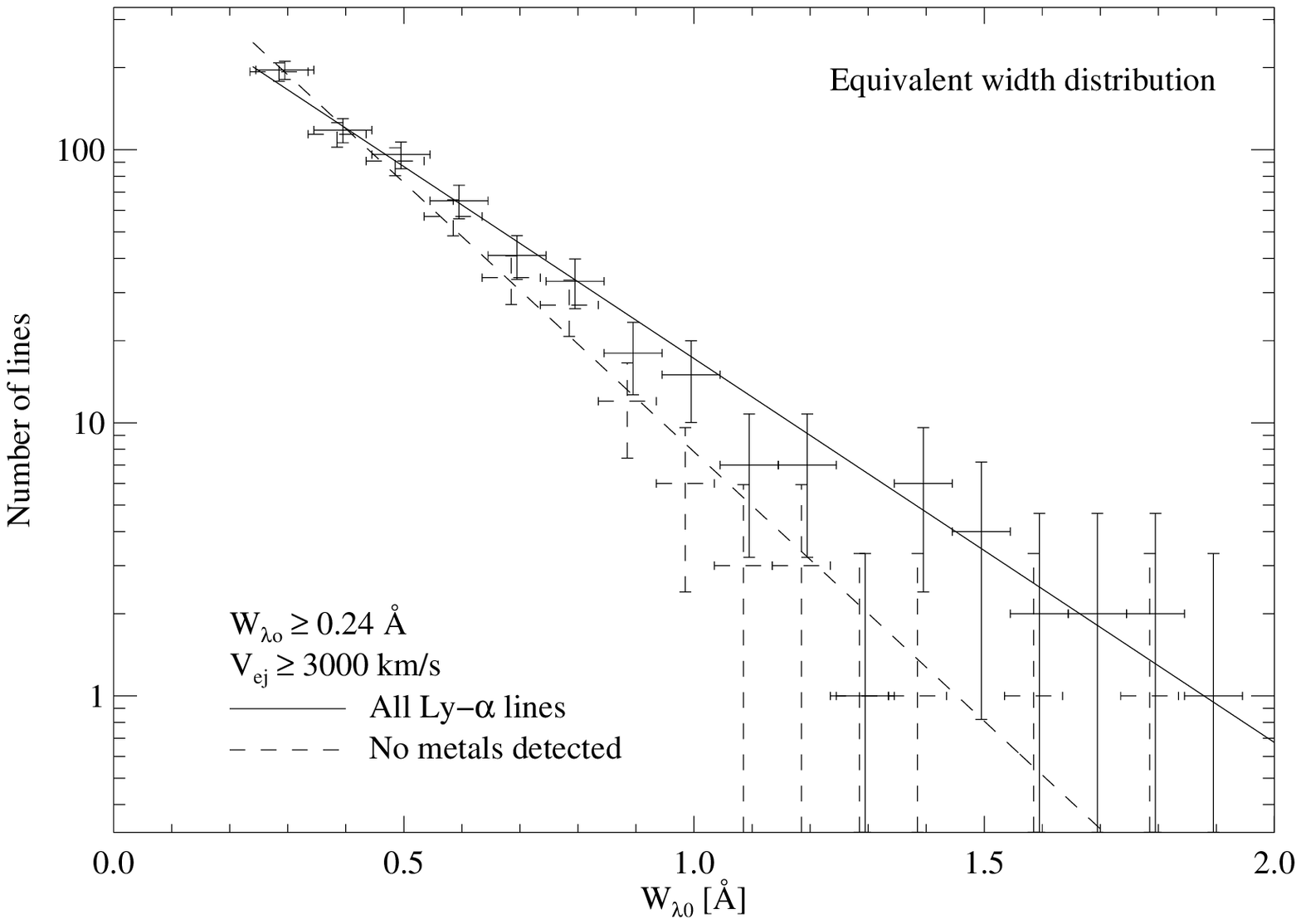}
\caption{Distribution of rest equivalent widths in the
$\Wrest\geq0.24$~\AA, $\Vej\geq3000$~\kms\ sample. The solid line
shows the data points and the fit of $d{\cal N}/d\Wrest \propto
\exp(-\Wrest /\Wstar)$ for all \Lya\ lines in the sample, the dashed
line shows the data and fit for the \Lya\ forest lines
only.\label{fig:wstarfit}}
\end{figure}

Equation~(\ref{eq:dndz}) describes very well the distribution of rest
equivalent widths of lines in the sample. We note, however, that the
derived value of the equivalent width distribution parameter, \Wstar,
depends quite strongly on whether \Lya\ lines associated with metal
systems are included in the analyzed sample. Figure~\ref{fig:wstarfit}
shows the distribution of equivalent widths in the $\Wrest \geq
0.24$~\AA, $\Vej\geq3000$~\kms\ samples. Clearly, inclusion of lines
from metal systems, which (as noted above) tend to be strong, affects
the slope of the observed distribution.

It has been found that a simple power law, $d{\cal N}/d\NHI \propto
\NHI^{-\beta}$, describes the distribution of \Lya\ forest column
densities \NHI. Analyses of high redshift data (Petitjean et al.\
1993; Press \& Rybicki 1993; Hu et al.\ 1995; Dobrzycki \& Bechtold
1996; Kim, Cristiani, \& D'Odorico 2001) showed that the value of
$\beta$ is in the $1.4-1.7$ range. Analysis of low redshift data from
HST/GHRS (Penton, Schull, \& Stocke 2000) indicated $\beta$ near the
high end of this range, at $1.7-1.8$. However, recent analysis by
Dav\'e \& Tripp (2001) of high resolution HST/STIS spectra of two
quasars suggested that at low $z$ the distribution is even steeper,
with $\beta\approx2.0$.

The FOS data lack the spectral resolution to perform direct Voigt
profile fitting, which is necessary to derive the column density
directly. However, because of the large size of the sample we can
obtain precise measurements of the equivalent width distribution.
Since column density and equivalent width are related via the curve of
growth (see, e.g., Barcons \& Webb 1991; Chernomordik \& Ozernoy 1993;
Penton, Schull, \& Stocke 2000), the distribution of line equivalent
widths can give insight into the distribution of column densities.

Our data do not support the $\beta\approx2$ value of the column
density distribution slope. For all our \Lya\ forest subsamples, we
get the equivalent width distribution parameter, \Wstar, in the
$0.20-0.22$~\AA\ range with very low uncertainties, of the order of
$0.01$~\AA\ or less. This value of \Wstar, when converted to the value
of $\beta$ using the curve of growth, yields $\beta \leq 1.6-1.7$,
depending on the Doppler parameter distribution. In order to obtain
the value of the slope in the vicinity of 2, we would have to have
\Wstar\ from the $0.10-0.13$~\AA\ range, which is inconsistent with
our results.

However, as Barcons \& Webb (1991) pointed out, the relation between
$\beta$ and $\Wstar$ may be affected by a subtle problem with line
blending, or the presence of clustering of weak lines. Since those
effects are hard to quantify, it is possible that the discrepancy is
in fact smaller. Also, as noted above, the derivation of \Wstar\ is
rather sensitive to the selection criteria in the sample.
Including/excluding lines associated with metal systems in the sample
leads to distinctly different value of the distribution slope. This
may help explain some of the discrepancy with the Dav\'e and Tripp
(2001) result.


\section{Clustering in the HST/FOS sample}
\label{sec-clustering}

An important clue to the nature of the \Lya\ absorption in quasar
spectra is the clustering of lines in redshift space. In general, it
is agreed that a weak clustering at scales of few hundred km/s is
present in high redshift \Lya\ forest (Rauch 1998 and references
therein; more recently Liske et al.\ 2000; Penton, Schull, \& Stocke
2000). At low redshifts, Ulmer (1996) analyzed the line clustering in
the spectra of 12 objects from the HST/FOS Key Project (100 lines) and
reported a significant excess of line pairs at velocity separations of
250--500~\kms. More recently, Vanden Berk et al.\ (1999) reported
positive correlation for similar velocity separations, though not as
strong as Ulmer, in a sample consisting of the combination of Ulmer's
sample and FOS spectra of quasars in the galactic poles (217
absorption lines, 22 quasars); they noted that the signal seems to be
coming from a small set of line ``clumps'' in a few of the objects.

Impey, Petry, \& Flint (1999) found excess of nearest-neighbor pairs
at separations of 250--500~\kms\ in their sample of 139 lines from 10
HST/GHRS spectra, but they did not find any correlation function
signal on any scales. However, Penton, Schull, \& Stocke (2000)
reported a clustering signal for $\Delta V \leq 150$~\kms\ in their
sample of 15 HST/GHRS spectra containing 111 lines.

The critical factor in the analysis of line clustering is the number
of the available objects (i.e.\ lines of sight), since correlating the
line positions is possible only in one dimension. Our sample is a
superset of the Ulmer sample and --- with the exception of two
quasars, TON155 and 1306+3021, which did not meet our sample's
criteria --- of the Vanden Berk et al.\ sample. In our data set that
has comparable selection criteria ($\Wrest \geq 0.24$~\AA, $\Vej \geq
3000$~\kms), we have 84 objects with 622 absorption lines, i.e.\ seven
times more lines of sight and six times more lines than Ulmer and ca.\
four times more lines of sight and three times more lines than Vanden
Berk et al.

\subsection{Lyman $\alpha$ -- Lyman $\alpha$ clustering}
\label{sec-lya}

We calculated the correlation function by comparing the observed
number of pairs of lines with the number counted in Monte Carlo
simulations. We performed the simulations by drawing redshifts at
random from the power-law distribution of $d{\cal N}/dz$
(Eq.~[\ref{eq:dndz}]) inside ranges identical to those in which our
sample was defined. Each simulation was run until we had a total
number of lines equal to the total number of lines in the real
sample. We ran the simulations 10,000 times. To account for blending,
we removed pairs with separations smaller than 250~\kms. However, we
do expect blending to play at least some role, so that the number of
expected pairs in the simulations may be overestimated, primarily in
the $250-500$~\kms\ bin.

\begin{figure}
\plotone{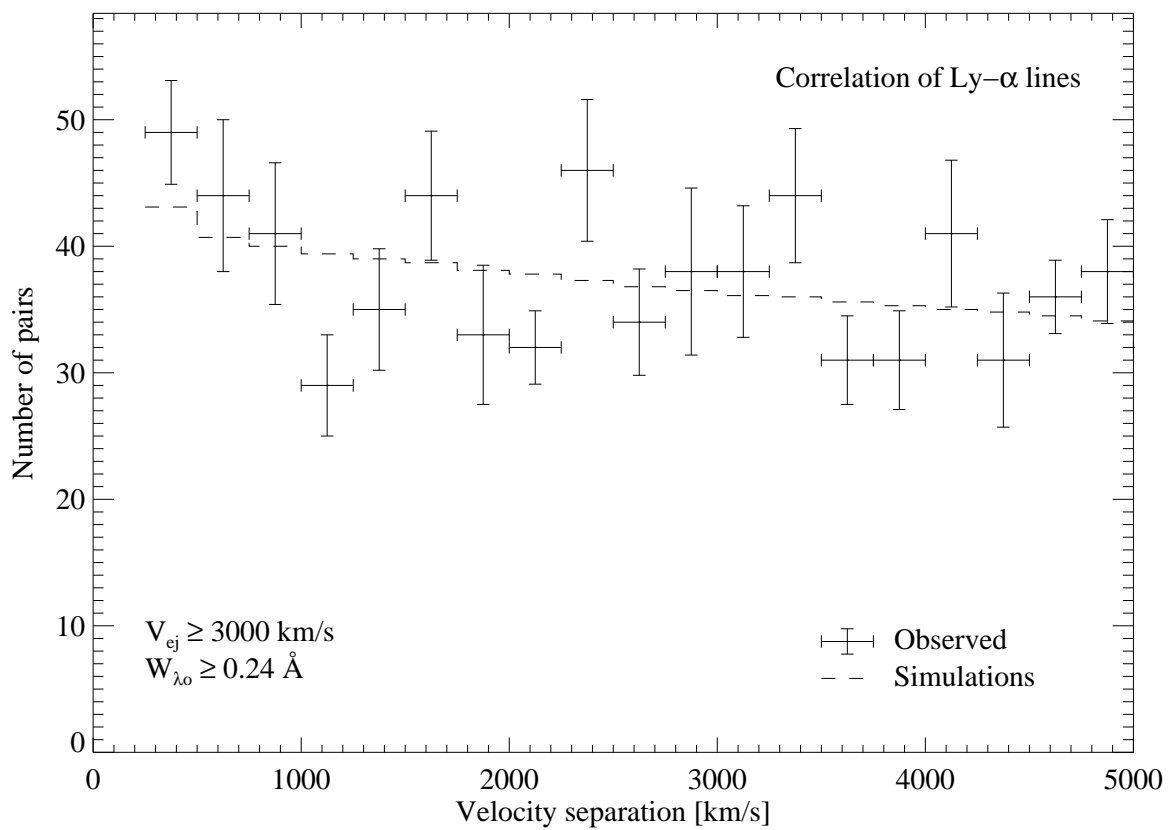}
\caption{Observed (solid crosses) and expected (dashed histogram)
number of pairs versus velocity separation in the $\Wrest \geq
0.24$~\AA, $\Vej\geq3000$~\kms\ sample, including the metal
systems. This plot is directly comparable to the lower panel of Fig.~1
of Ulmer (1996). Error bars were estimated using bootstrap technique,
see text.\label{fig:corr}}
\end{figure}

The clustering results in our sample are shown in
Figure~\ref{fig:corr}. Solid crosses show the observed number of pairs
in each velocity separation bin, while the dashed histogram shows the
expected number of pairs, as calculated through the simulations.

A key issue is the uncertainties of the observed data points. If the
lines are clustered, simply taking the square root of the number of
line pairs in each bin will cause the error of the bin to be
underestimated. We calculated the error bars for the observed data
points using a modified bootstrap technique, similar to the method
used by Fern\'andez-Soto (1996). In this method, one replaces parts of
the spectra with randomly selected substitute parts of the spectra and
then calculates the distribution of pairs, repeating this procedure a
large number of times. It can be shown that the dispersion of pair
counts in each bin is an unbiased estimate of the actual variance in
this bin.

In Fig.~\ref{fig:corr}, a weak correlation can be seen for $\Delta
V<500$~\kms. The signal is weak, though --- as mentioned above --- it
is quite possible that line blending was not all accounted for in the
simulations, leading to overestimating the expected number of pairs,
thus mitigating the clustering signal.

The correlation found in our data has lower amplitude than the one
reported by Ulmer (1996) or Vanden Berk et al.\ (1999). At least part
of the difference can be explained if one looks into the sample used
by Ulmer (and, consequently, Vanden Berk et al.) in more detail.

\begin{figure}
\plotone{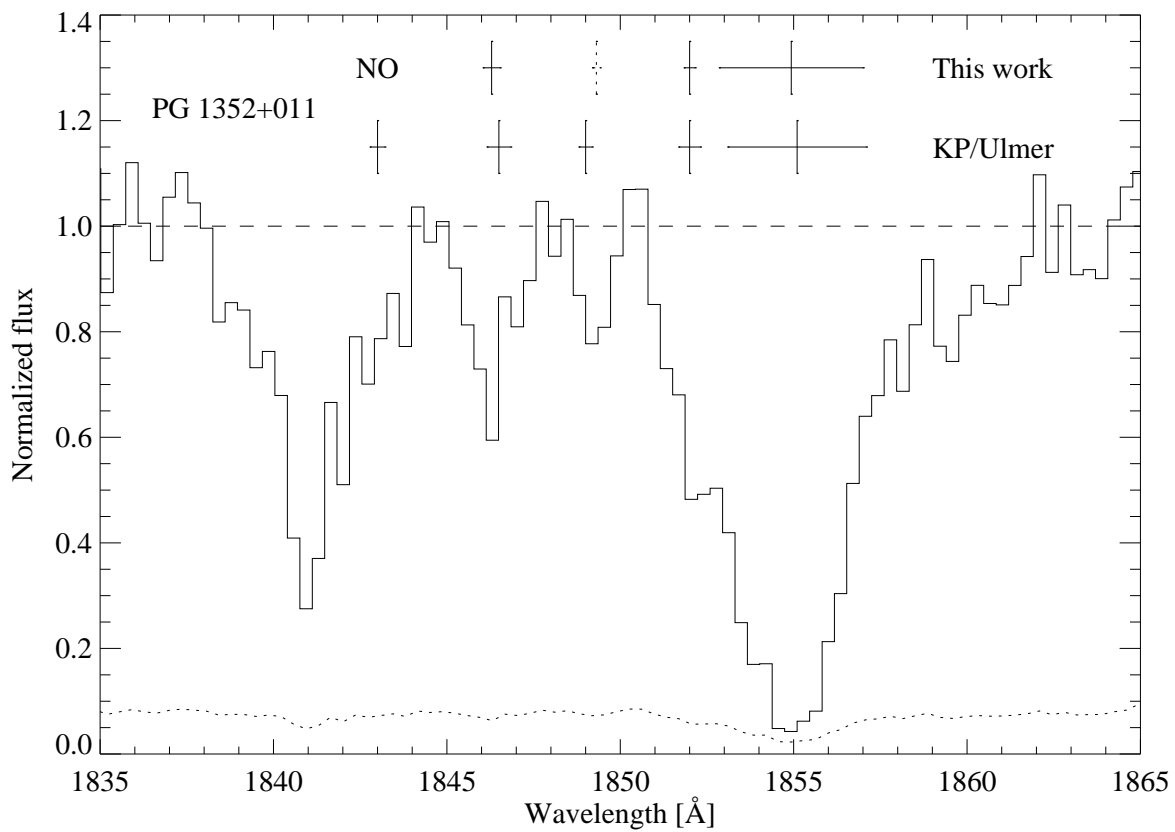}
\caption{Region of the spectrum of PG~1352+011 where a difference
between our analysis (Paper~III) and the Key Project leads to a
difference in the clustering signal. Dotted line shows 1-$\sigma$
error in the spectrum. Upper row of crosses above absorption features
represent our identifications, the lower crosses represent those of
the Key Project. The dotted symbol denotes a line identified in
Paper~III, but at a significance too low to be included in our
analysis. ``NO'' marks a location of a Key Project line that is not
identified as a separate feature in our analysis. See text for
discussion.\label{fig:kp2us1}}
\end{figure}

We first note that our selection criteria excluded regions containing
the cluster of four pairs in TON~153, since this region is bluewards
of the \Lyb\ emission line. In regions that we have in common, we note
the following differences:

1. In PG~1352+011, Ulmer has four close pairs, coming from lines near
1843, 1846, 1849, 1852, and 1855~\AA. We present this region in
Figure~\ref{fig:kp2us1}. The line at 1843~\AA\ was deblended by Key
Project from a wing of a strong Si\thinspace{\sc iii} line near
1841~\AA. In our analysis (Paper~III) it was not identified as a
separate feature. We do see a line near 1849~\AA\ (Paper~III), but it
has a detection significance of $2.7\sigma$, which is too low to be
included in the sample analyzed here. All these factors result in our
sample only containing one close pair in this region, the 1852--1855
pair.

2. In two pairs in Ulmer's sample --- $\lambda=1562.7$ in 3C~351 and
$\lambda=2601.6$ in PG~1634+706 --- we identify one of the pair
components as a heavy element line rather than as a \Lya\ line. This
means that they are removed from consideration in our analysis of
correlations. Both these systems were considered by Ulmer (and Vanden
Berk et al.) to be \Lya\ pairs.

In total, in the regions that we had in common, we do see five fewer
close pairs. Removal of those pairs lowers the number reported by
Ulmer (14 pairs) by 36\%, and the correlation signal (i.e.\ the ratio
of $N_{\rm obs}/N_{\rm ran}$) by about half. Indeed, as it has been
found in previous studies (Heisler, Hogan, \& White 1989; Bahcall et
al.\ 1996; Vanden Berk et al.\ 1999), the correlation signal is
dependent on a small handfull of line pairs.

\begin{figure}
\plotone{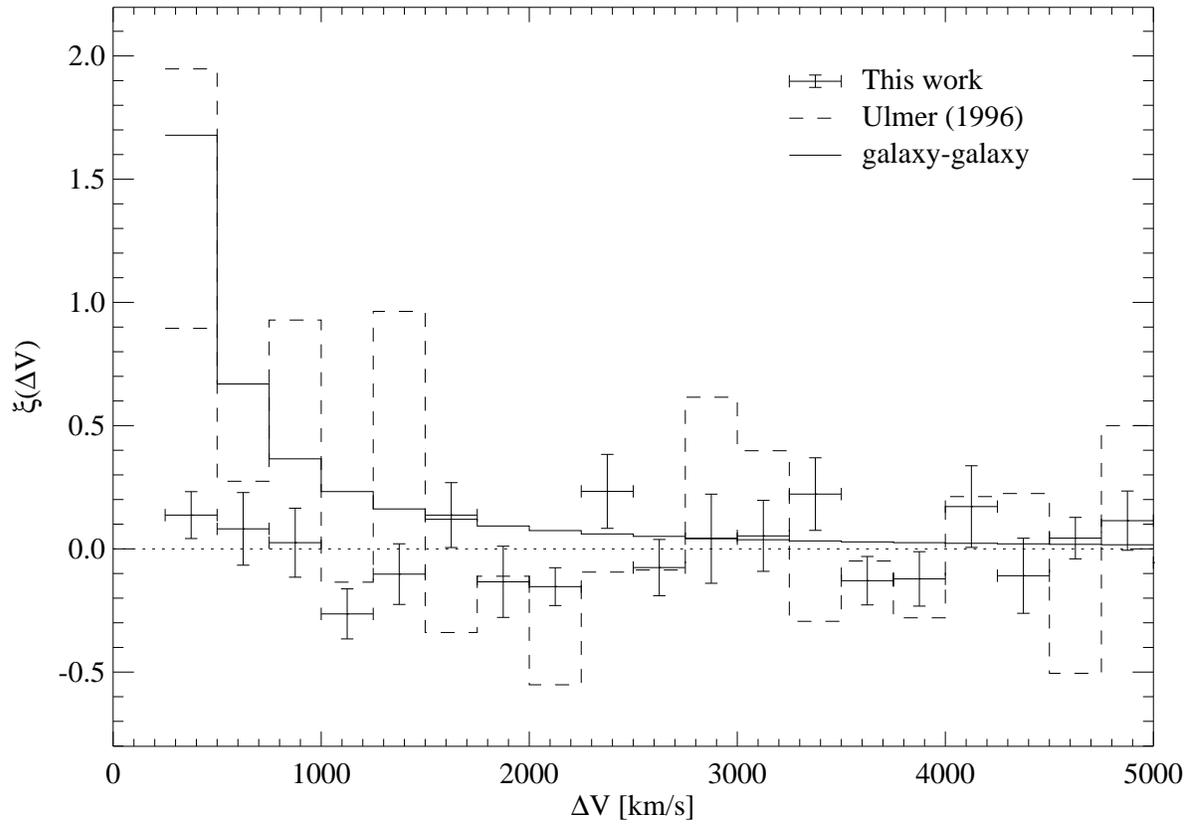}
\caption{Two point correlation function, $\xi(\Delta V)$, for our
data (solid crosses), Ulmer (1996) (dashed line), and galaxy-galaxy
correlation (solid line). Note two data points for Ulmer's data in the
$250-500$~\kms\ bin; the lower point indicates the correlation
function one gets if one removes five disputable pairs from Ulmer's
sample.\label{fig:corrfun}}
\end{figure}

In Fig.~\ref{fig:corrfun} we show the results converted to the two
point correlation function, $\xi(\Delta V) = N_{\rm obs}/N_{\rm
ran}-1$. On the same plot, the dashed line shows Ulmer's result. We
show two data points for Ulmer's results in the $250-500$~\kms\ bin;
the bottom one is what his result would be if the five close pairs
mentioned above were removed. Please note that we examined in detail
only the $250-500$~\kms\ bin in Ulmer's data and so we do not show how
other velocity bins would be affected. Obviously, the differences in
line identifications and deblending accounts for some --- but not all
--- of the difference between the correlation strengths.

Clustering of \Lya\ lines at small velocity separations is predicted
by Cold Dark Matter (CDM) model simulations of the galaxy formation,
the growth of large scale structure and subsequent evolution of the
intergalactic medium (Cen et al.\ 1998; Dav\'e et al.\ 1999). These
models predict that the \Lya\ absorbing gas traces the dark matter
distribution more closely than galaxies, and so the 2-point
correlation function of \Lya\ clouds provides a different and
complementary constraint on the formation of large scale structure
than the distribution of galaxies and galaxy clusters. The solid
histogram on Fig.~\ref{fig:corrfun} shows the galaxy-galaxy
correlation function, taken as $\xi_{\rm g-g} = (\Delta V/500)^{-1.8}$
(e.g.\ Davis \& Peebles 1983). We see that the galaxy-galaxy
correlation function is stronger than the \Lya\ correlation function,
as is predicted by the CDM simulations.

\subsection{Lyman $\alpha$ -- metal system correlation}
\label{sec-metals}

Analyses of FOS data suggested that \Lya\ forest lines may have a
tendency to be found preferably near heavy element systems. This
tendency appeared to be even stronger if one considered the
``extensive'' metal systems, i.e.\ metal systems with four or more
identified species; those systems were often seen near ``clumps'' of
\Lya\ lines. See Bahcall et al.\ (1996), Jannuzi (1997), and Vanden
Berk et al.\ (1999). Given scarcity of available systems, those
analyses were, for the most part, qualitative, though Vanden Berk et
al.\ reported some positive signal for metal system--\Lya\ forest
correlation at $\Delta V = 1000-1500$~\kms\ for strong
($\Wrest>0.5$~\AA) lines.

\begin{figure}
\plotone{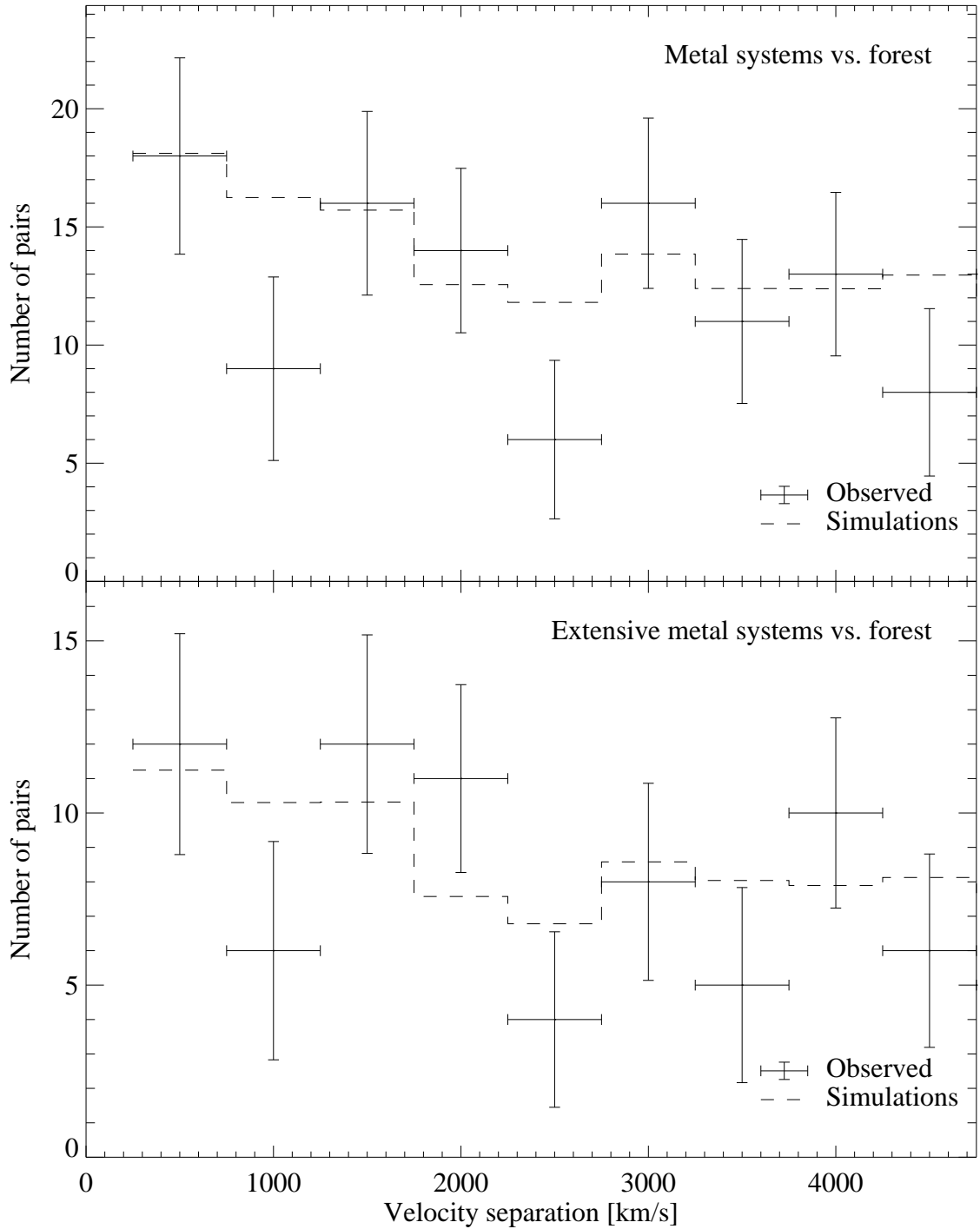}
\caption{Distribution of line velocity separations, as in
Figure~\ref{fig:corr}, between metal systems and Lyman alpha forest
lines. Top panel: \Lya\ forest lines vs.\ metal line systems. Bottom
panel: \Lya\ forest vs.\ extensive metal
systems.\label{fig:corrmetals}}
\end{figure}

In the $\Wrest>0.24$~\AA, $\Vej\geq3000$~\kms\ sample, we have 78
identified metal systems, 41 of which qualify as ``extensive''. We
analyzed the correlation of \Lya\ forest vs.\ those systems, using
approach similar to the one used in the analysis of \Lya\ forest
correlations. In the simulations, we fixed the positions of the metal
systems at their observed locations and selected locations of \Lya\
forest lines at random. We compared the simulated and observed pairs
of absorbers, requiring that one of the pair components be a metal
system. We do not see any correlation on any scale in both cases
(Fig.~\ref{fig:corrmetals}). It is, however, clear that even the
entire FOS sample is somewhat too small for this test.


\section{Summary}
\label{sec-summary}

We analyzed the redshift distribution of the \Lya\ clouds at
$z=0.0-1.7$ in the spectra of 165 quasars. The major results are the
following.

1. We confirm that the clouds at redshifts less than 1.7 undergo slow,
if any, evolution. A single power law is an excellent approximation of
the behavior of the number density evolution of lines at low
redshift. However, we find the evolution of the cloud number density
to be somewhat steeper that the one found by the HST Key Project,
probably the result of a different proportion of $z\sim1.5-1.7$ lines
in the two samples.

2. We see marginal evidence for a difference between the evolution
index for weak and strong lines. The weak lines undergo evolution in
the opposite direction from the strong lines, i.e.\ their number per
comoving volume increases with decreasing redshift. The strong lines
are consistent with the no-evolution case.

3. The object-to-object variation in the number of lines is small,
suggesting that the intergalactic medium is uniform on $\sim$800~Mpc
scales.

4. We see weak clustering in the \Lya\ forest for $\Delta V \leq
500$~\kms. However, the correlation is weaker than reported by Ulmer
(1996) or Vanden Berk et al.\ (1999). At least part of the difference
can be accounted for by differences in details of the deblending and
identification of a small number of absorption features. In all cases,
the clustering signal is smaller than that seen for galaxies at the
same redshift.

5. No correlations were found for \Lya\ forest vs.\ metal systems and
\Lya\ forest vs.\ ``extensive'' metal systems, although the numbers
are small, and the limits correspondedly are weak.

The FOS data set is useful for many other quasar absorption line
studies. Paper~V reports the analysis of the proximity effect, and the
evolution of the UV radiation field which may drive the evolution of
$d{\cal N}/dz$ for the \Lya\ forest clouds derived here (see Paper~V
and references therein for further discussion of this point). The
results for clustering described in this paper can be confirmed or
refuted with high resolution spectroscopy, currently planned or
underway with STIS on board HST.


\acknowledgements

AD would like to thank A. Fern\'andez-Soto for helpful comments on the
bootstrapping technique. We would like to thank the anonymous referee
for several helpful comments. This research has made use of the
NASA/IPAC Extragalactic Database (NED) which is operated by the Jet
Propulsion Laboratory, California Institute of Technology, under
contract with NASA. This project was supported by STScI grants No.\
AR-05785.02-94A and GO-06606.01-95. AD acknowledges support from NASA
Contract No.\ NAS8-39073 (CXC). JB, JS and MM received financial
support from NSF grant AST-9617060. JS acknowledges support of the
National Science Foundation Graduate Research Fellowship and the Zonta
Foundation Amelia Earhart Fellowship.



\clearpage

\clearpage

\clearpage

\clearpage

\clearpage

\clearpage

\clearpage

\clearpage

\clearpage

\clearpage

\clearpage

\clearpage

\begin{deluxetable}{lccclccc}
\tabletypesize{\scriptsize}
\tablewidth{0pt}
\tablecaption{HST/FOS quasar sample.\label{tab:quasars}}
\tablehead{%
\colhead{Designation} &
\colhead{$z$} &
\colhead{$\alpha_{\rm 1950}$} &
\colhead{$\delta_{\rm 1950}$} &
\colhead{Object name} &
\colhead{$z_{\rm beg}$\tablenotemark{a}} &
\colhead{$z_{\rm end}$\tablenotemark{b}} &
\colhead{U/V/W\tablenotemark{c}}
}
\startdata
Q0002$+$0507   &  1.900 &  00 02 46.4 &  $+$05 07 29  & UM18                     & 1.447 & 1.693 &  W \\
Q0003$+$1553   &  0.450 &  00 03 25.1 &  $+$15 53 07  & 0003$+$15                & 0.223 & 0.450 &  W \\
Q0003$+$1955   &  0.025 &  00 03 45.3 &  $+$19 55 28  & MRK335                   & 0.006 & 0.025 &  \\
Q0007$+$1041   &  0.089 &  00 07 56.7 &  $+$10 41 47  & IIIZW2                   & 0.007 & 0.089 &  \\
Q0015$+$1612   &  0.553 &  00 15 56.7 &  $+$16 12 46  & QSO0015$+$162            & 0.310 & 0.553 &  \\
Q0017$+$0209   &  0.401 &  00 17 51.1 &  $+$02 09 46  & Q0017$+$0209             & 0.330 & 0.401 &  \\
Q0024$+$2225   &  1.108 &  00 24 38.6 &  $+$22 25 23  & NAB0024$+$22             & 0.779 & 1.108 &  W \\
Q0026$+$1259   &  0.142 &  00 26 38.1 &  $+$12 59 29  & PG0026$+$12              & 0.001 & 0.142 &  \\
Q0042$+$1010   &  0.583 &  00 42 22.8 &  $+$10 10 28  & MC0042$+$101             & 0.368 & 0.583 &  \\
Q0043$+$0354   &  0.384 &  00 43 12.6 &  $+$03 54 00  & PG0043$+$039             & 0.354 & 0.384 &  W \\
Q0044$+$0303   &  0.624 &  00 44 31.5 &  $+$03 03 32  & PKS0044$+$030            & 0.370 & 0.624 &  W \\
Q0050$+$1225   &  0.061 &  00 50 58.0 &  $+$12 25 20  & IZW1                     & 0.007 & 0.061 &  \\
Q0100$+$0205   &  0.394 &  01 00 38.6 &  $+$02 05 04  & 0100$+$0205              & 0.295 & 0.394 &  \\
Q0102$-$2713   &  0.780 &  01 02 16.6 &  $-$27 13 12  & CT336                    & 0.502 & 0.780 &  V \\
Q0107$-$0235   &  0.948 &  01 07 40.3 &  $-$02 35 51  & Q0107$-$025A             & 0.644 & 0.899 &  \\
Q0107$-$1537   &  0.861 &  01 07 03.2 &  $-$15 37 50  & QSO0107$-$156            & 0.831 & 0.861 &  \\
Q0117$+$2118   &  1.493 &  01 17 34.7 &  $+$21 18 02  & PG0117$+$213             & 1.103 & 1.493 &  W \\
Q0121$-$5903   &  0.047 &  01 21 51.2 &  $-$59 03 59  & FAIRALL9                 & 0.013 & 0.047 &  \\
Q0122$-$0021   &  1.070 &  01 22 55.3 &  $-$00 21 31  & PKS0122$-$003            & 0.747 & 1.070 &  UW \\
Q0137$+$0116   &  0.260 &  01 37 22.9 &  $+$01 16 35  & PHL1093                  & 0.234 & 0.260 &  \\
Q0159$-$1147   &  0.669 &  01 59 30.4 &  $-$11 47 00  & 3C57                     & 0.408 & 0.669 &  W \\
Q0214$+$1050   &  0.408 &  02 14 26.8 &  $+$10 50 18  & PKS0214$+$10             & 0.295 & 0.408 &  \\
Q0219$+$4248   &  0.444 &  02 19 30.1 &  $+$42 48 30  & 3C66A                    & 0.218 & 0.444 &  \\
Q0232$-$0415   &  1.450 &  02 32 36.6 &  $-$04 15 10  & PKS0232$-$04             & 1.067 & 1.450 &  W \\
Q0254$-$3327B  &  1.915 &  02 54 39.4 &  $-$33 27 24  & PKS0254$-$334            & 1.460 & 1.693 &  \\
Q0302$-$2223   &  1.400 &  03 02 35.7 &  $-$22 23 29  & 1E0302$-$223             & 1.025 & 1.397 &  \\
Q0333$+$3208   &  1.258 &  03 33 22.5 &  $+$32 08 36  & NRAO140                  & 0.905 & 1.251 &  \\
Q0349$-$1438   &  0.616 &  03 49 09.6 &  $-$14 38 06  & 3C95                     & 0.364 & 0.616 &  UW \\
Q0355$-$4820   &  1.005 &  03 55 52.5 &  $-$48 20 48  & 0355$-$4820              & 0.828 & 1.005 &  \\
Q0405$-$1219   &  0.574 &  04 05 27.5 &  $-$12 19 31  & PKS0405$-$12             & 0.328 & 0.574 &  W \\
Q0414$-$0601   &  0.781 &  04 14 49.3 &  $-$06 01 05  & PKS0414$-$06             & 0.503 & 0.781 &  W \\
Q0420$-$0127   &  0.915 &  04 20 43.5 &  $-$01 27 29  & PKS0420$-$01             & 0.830 & 0.915 &  \\
Q0421$+$0157   &  2.044 &  04 21 32.7 &  $+$01 57 32  & PKS0421$+$01             & 1.568 & 1.685 &  \\
Q0424$-$1309   &  2.159 &  04 24 47.7 &  $-$13 09 33  & PKS0424$-$13             & 1.678 & 1.693 &  \\
Q0439$-$4319   &  0.593 &  04 39 42.8 &  $-$43 19 24  & PKS0439$-$433            & 0.344 & 0.593 &  W \\
Q0454$-$2203   &  0.534 &  04 54 01.2 &  $-$22 03 49  & PKS0454$-$22             & 0.294 & 0.534 &  \\
Q0454$+$0356   &  1.345 &  04 54 09.0 &  $+$03 56 14  & PKS0454$+$039            & 0.979 & 1.345 &  \\
Q0518$-$4549   &  0.035 &  05 18 23.6 &  $-$45 49 43  & PKS0518$-$45             & 0.002 & 0.035 &  \\
Q0624$+$6907   &  0.374 &  06 24 35.3 &  $+$69 07 03  & HS0624$+$6907            & 0.159 & 0.374 &  W \\
Q0637$-$7513   &  0.656 &  06 37 23.5 &  $-$75 13 37  & PKS0637$-$75             & 0.397 & 0.656 &  W \\
Q0742$+$3150   &  0.462 &  07 42 30.8 &  $+$31 50 15  & B20742$+$318             & 0.329 & 0.462 &  W \\
Q0743$-$6719   &  1.511 &  07 43 22.3 &  $-$67 19 07  & PKS0743$-$67             & 1.119 & 1.511 &  W \\
Q0823$-$2220   &  0.910 &  08 23 50.0 &  $-$22 20 34  & PKS0823$-$22             & 0.828 & 0.910 &  \\
Q0827$+$2421   &  0.935 &  08 27 54.4 &  $+$24 21 07  & B20827$+$24              & 0.831 & 0.935 &  \\
Q0844$+$3456   &  0.064 &  08 44 34.0 &  $+$34 56 07  & TON951                   & 0.007 & 0.064 &  \\
Q0848$+$1623   &  1.936 &  08 48 53.7 &  $+$16 23 40  & Q0848$+$163              & 1.479 & 1.694 &  \\
Q0850$+$4400   &  0.513 &  08 50 13.5 &  $+$44 00 24  & US1867                   & 0.327 & 0.513 &  W \\
Q0916$+$5118   &  0.553 &  09 16 30.1 &  $+$51 18 52  & NGC2841UB3               & 0.310 & 0.553 &  W \\
Q0935$+$4141   &  1.937 &  09 35 48.8 &  $+$41 41 55  & PG0935$+$416             & 1.478 & 1.693 &  \\
Q0945$+$4053   &  1.252 &  09 45 50.1 &  $+$40 53 43  & 4C40$-$24                & 0.900 & 1.252 &  \\
Q0947$+$3940   &  0.206 &  09 47 44.9 &  $+$39 40 54  & PG0947$+$396             & 0.018 & 0.206 &  \\
Q0953$+$4129   &  0.239 &  09 53 48.3 &  $+$41 29 40  & PG0953$+$414             & 0.045 & 0.239 &  W \\
Q0955$+$3238   &  0.533 &  09 55 25.5 &  $+$32 38 23  & 3C232                    & 0.293 & 0.533 &  W \\
Q0957$+$5608A  &  1.414 &  09 57 57.4 &  $+$56 08 22  & 0957$+$561A              & 1.041 & 1.414 &  \\
Q0958$+$5509   &  1.750 &  09 58 08.2 &  $+$55 09 06  & MARK132                  & 1.327 & 1.694 &  \\
Q0959$+$6827   &  0.773 &  09 59 09.7 &  $+$68 27 47  & 0959$+$68W1              & 0.496 & 0.773 &  W \\
Q1001$+$0527   &  0.161 &  10 01 43.3 &  $+$05 27 34  & PG1001$+$054             & 0.002 & 0.161 &  \\
Q1001$+$2239   &  0.974 &  10 01 58.5 &  $+$22 39 54  & PKS1001$+$22             & 0.830 & 0.974 &  \\
Q1001$+$2910   &  0.329 &  10 01 10.7 &  $+$29 10 08  & TON28                    & 0.121 & 0.329 &  W \\
Q1007$+$4147   &  0.611 &  10 07 26.1 &  $+$41 47 25  & 4C41.21                  & 0.359 & 0.611 &  W \\
Q1008$+$1319   &  1.287 &  10 08 29.9 &  $+$13 19 00  & PG1008$+$133             & 0.930 & 1.287 &  W \\
Q1010$+$3606   &  0.070 &  10 10 07.4 &  $+$36 06 15  & CSO251                   & 0.004 & 0.070 &  \\
Q1017$+$2759   &  1.928 &  10 17 07.8 &  $+$27 59 06  & TON34                    & 1.470 & 1.693 &  \\
Q1026$-$0045A  &  1.437 &  10 26 01.7 &  $-$00 45 22  & Q1026$-$0045$-$A         & 1.056 & 1.437 &  \\
Q1026$-$0045B  &  1.530 &  10 26 03.6 &  $-$00 45 06  & Q1026$-$0045$-$B         & 1.141 & 1.530 &  \\
Q1038$+$0625   &  1.270 &  10 38 41.0 &  $+$06 25 58  & 4C06.41                  & 0.915 & 1.270 &  W \\
Q1049$-$0035   &  0.357 &  10 49 18.1 &  $-$00 35 21  & PG1049$-$005             & 0.327 & 0.357 &  W \\
Q1055$+$2007   &  1.110 &  10 55 37.6 &  $+$20 07 55  & PKS1055$+$20             & 0.830 & 1.110 &  \\
Q1100$+$7715   &  0.311 &  11 00 27.5 &  $+$77 15 08  & 3C249$-$1                & 0.106 & 0.311 &  W \\
Q1104$+$1644   &  0.634 &  11 04 36.7 &  $+$16 44 16  & MC1104$+$167             & 0.379 & 0.634 &  W \\
Q1114$+$4429   &  0.144 &  11 14 20.0 &  $+$44 29 57  & PG1114$+$445             & 0.002 & 0.144 &  \\
Q1115$+$0802A1 &  1.722 &  11 15 41.5 &  $+$08 02 23  & PG1115$+$080A1           & 1.309 & 1.694 &  \\
Q1115$+$4042   &  0.154 &  11 15 45.9 &  $+$40 42 19  & PG1115$+$407             & 0.002 & 0.154 &  \\
Q1116$+$2135   &  0.177 &  11 16 30.2 &  $+$21 35 43  & PG1116$+$215             & 0.004 & 0.177 &  W \\
Q1118$+$1252   &  0.685 &  11 18 53.5 &  $+$12 52 43  & MC1118$+$12              & 0.422 & 0.685 &  \\
Q1130$+$1108   &  0.510 &  11 30 55.0 &  $+$11 08 57  & 1130$+$106Y              & 0.337 & 0.510 &  W \\
Q1136$-$1334   &  0.557 &  11 36 38.6 &  $-$13 34 05  & PKS1136$-$135            & 0.334 & 0.557 &  W \\
Q1137$+$6604   &  0.652 &  11 37 09.4 &  $+$66 04 27  & 3C263.0                  & 0.394 & 0.652 &  UW \\
Q1138$+$0204   &  0.383 &  11 38 47.8 &  $+$02 04 41  & Q1138$+$0204             & 0.351 & 0.383 &  \\
Q1146$+$1104B  &  1.010 &  11 46 09.8 &  $+$11 04 37  & 1146$+$111B              & 0.696 & 0.899 &  \\
Q1146$+$1106C  &  1.010 &  11 46 04.9 &  $+$11 06 57  & 1146$+$111C              & 0.696 & 0.891 &  \\
Q1148$+$5454   &  0.969 &  11 48 42.6 &  $+$54 54 13  & 1148$+$5454              & 0.661 & 0.969 &  \\
Q1150$+$4947   &  0.334 &  11 50 48.1 &  $+$49 47 50  & LB2136                   & 0.295 & 0.334 &  \\
Q1156$+$2123   &  0.349 &  11 56 52.3 &  $+$21 23 38  & TEX1156$+$213            & 0.296 & 0.349 &  \\
Q1156$+$2931   &  0.729 &  11 56 57.9 &  $+$29 31 25  & 4C29.45                  & 0.459 & 0.729 &  \\
Q1206$+$4557   &  1.158 &  12 06 26.6 &  $+$45 57 17  & PG1206$+$459             & 0.821 & 1.158 &  \\
Q1211$+$1419   &  0.085 &  12 11 45.0 &  $+$14 19 52  & PG1211$+$1431            & 0.012 & 0.085 &  \\
Q1214$+$1804   &  0.375 &  12 14 16.8 &  $+$18 04 44  & Q1214$+$1804             & 0.344 & 0.373 &  \\
Q1215$+$6423   &  1.288 &  12 15 17.1 &  $+$64 23 46  & 4C64$-$15                & 0.931 & 1.288 &  \\
Q1216$+$0655   &  0.334 &  12 16 47.8 &  $+$06 55 17  & PG1216$+$069             & 0.126 & 0.334 &  W \\
Q1216$+$5032A  &  1.450 &  12 16 13.5 &  $+$50 32 15  & HS1216$+$5032A           & 1.067 & 1.450 &  \\
Q1219$+$0447   &  0.094 &  12 19 04.7 &  $+$04 47 03  & 1219$+$047               & 0.006 & 0.094 &  \\
Q1226$+$0219   &  0.158 &  12 26 33.2 &  $+$02 19 42  & 3C273                    & 0.002 & 0.158 &  UW \\
Q1229$-$0207   &  1.045 &  12 29 26.0 &  $-$02 07 32  & PKS1229$-$02             & 0.725 & 1.045 &  \\
Q1241$+$1737   &  1.273 &  12 41 41.0 &  $+$17 37 28  & PG1241$+$176             & 0.918 & 1.273 &  W \\
Q1247$+$2647   &  2.043 &  12 47 39.1 &  $+$26 47 26  & PG1247$+$267             & 1.568 & 1.694 &  \\
Q1248$+$3032   &  1.061 &  12 48 00.2 &  $+$30 32 58  & B21248$+$30              & 0.831 & 1.061 &  V \\
Q1248$+$3142   &  1.020 &  12 48 25.4 &  $+$31 42 11  & CSO173                   & 0.704 & 0.899 &  V \\
Q1248$+$4007   &  1.030 &  12 48 26.6 &  $+$40 07 58  & PG1248$+$401             & 0.713 & 1.030 &  W \\
Q1249$+$2929   &  0.820 &  12 49 59.6 &  $+$29 29 38  & CSO176                   & 0.536 & 0.820 &  V \\
Q1250$+$3122   &  0.780 &  12 50 52.9 &  $+$31 22 06  & CSO179                   & 0.502 & 0.780 &  V \\
Q1252$+$1157   &  0.871 &  12 52 07.7 &  $+$11 57 21  & PKS1252$+$11             & 0.579 & 0.871 &  W \\
Q1257$+$3439   &  1.375 &  12 57 26.6 &  $+$34 39 31  & B2011257$+$34            & 1.004 & 1.375 &  VW \\
Q1258$+$2835   &  1.355 &  12 58 36.6 &  $+$28 35 52  & QSO1258$+$285            & 0.987 & 1.355 &  \\
Q1259$+$5918   &  0.472 &  12 59 08.3 &  $+$59 18 14  & PG1259$+$593             & 0.242 & 0.472 &  UW \\
Q1302$-$1017   &  0.286 &  13 02 55.9 &  $-$10 17 17  & PKS1302$-$102            & 0.085 & 0.286 &  W \\
Q1305$+$0658   &  0.602 &  13 05 22.6 &  $+$06 58 14  & 3C281                    & 0.352 & 0.602 &  \\
Q1309$+$3531   &  0.184 &  13 09 58.4 &  $+$35 31 15  & PG1309$+$355             & 0.007 & 0.184 &  V \\
Q1317$+$2743   &  1.022 &  13 17 34.4 &  $+$27 43 51  & TON153                   & 0.706 & 1.022 &  UVW \\
Q1318$+$2903   &  0.549 &  13 18 54.7 &  $+$29 03 01  & TON156                   & 0.319 & 0.549 &  V \\
Q1320$+$2925   &  0.960 &  13 20 59.9 &  $+$29 25 45  & TON157                   & 0.654 & 0.958 &  V \\
Q1322$+$6557   &  0.168 &  13 22 08.5 &  $+$65 57 24  & PG1322$+$659             & 0.002 & 0.168 &  \\
Q1323$+$6530   &  1.618 &  13 23 48.6 &  $+$65 30 47  & 4C65.15                  & 1.209 & 1.618 &  \\
Q1327$-$2040   &  1.169 &  13 27 24.3 &  $-$20 40 48  & PKS1327$-$206            & 0.831 & 1.169 &  \\
Q1328$+$3045   &  0.849 &  13 28 49.7 &  $+$30 45 58  & 3C286.0                  & 0.560 & 0.849 &  V \\
Q1329$+$4117   &  1.930 &  13 29 29.8 &  $+$41 17 23  & PG1329$+$412             & 1.472 & 1.694 &  \\
Q1333$+$1740   &  0.554 &  13 33 36.8 &  $+$17 40 30  & PG1333$+$176             & 0.347 & 0.554 &  W \\
Q1338$+$4138   &  1.219 &  13 38 52.1 &  $+$41 38 22  & PG1338$+$416             & 0.872 & 1.213 &  W \\
Q1351$+$3153   &  1.326 &  13 51 51.3 &  $+$31 53 45  & B21351$+$31              & 0.963 & 1.326 &  \\
Q1351$+$6400   &  0.088 &  13 51 46.4 &  $+$64 00 29  & PG1351$+$64              & 0.008 & 0.088 &  \\
Q1352$+$0106   &  1.117 &  13 52 25.6 &  $+$01 06 51  & PG1352$+$011             & 0.790 & 1.117 &  UW \\
Q1352$+$1819   &  0.152 &  13 52 12.6 &  $+$18 19 58  & PG1352$+$183             & 0.002 & 0.152 &  \\
Q1354$+$1933   &  0.719 &  13 54 42.2 &  $+$19 33 43  & PKS1354$+$19             & 0.450 & 0.719 &  W \\
Q1404$+$2238   &  0.098 &  14 04 02.5 &  $+$22 38 03  & PG1404$+$226             & 0.002 & 0.098 &  \\
Q1407$+$2632   &  0.944 &  14 07 07.8 &  $+$26 32 30  & PG1407$+$265             & 0.640 & 0.944 &  W \\
Q1415$+$4509   &  0.114 &  14 15 04.7 &  $+$45 09 56  & PG1415$+$451             & 0.002 & 0.114 &  \\
Q1416$+$0642   &  1.436 &  14 16 38.8 &  $+$06 42 20  & 3C298                    & 1.055 & 1.436 &  \\
Q1424$-$1150   &  0.806 &  14 24 56.0 &  $-$11 50 25  & PKS1424$-$118            & 0.524 & 0.806 &  W \\
Q1425$+$2645   &  0.366 &  14 25 21.9 &  $+$26 45 39  & B21425$+$26              & 0.295 & 0.366 &  \\
Q1427$+$4800   &  0.221 &  14 27 54.0 &  $+$48 00 44  & PG1427$+$480             & 0.030 & 0.221 &  \\
Q1435$-$0134   &  1.310 &  14 35 13.3 &  $-$01 34 13  & Q1435$-$0134             & 0.949 & 1.304 &  \\
Q1435$+$6349   &  2.068 &  14 35 37.3 &  $+$63 49 36  & S41435$+$638             & 1.589 & 1.693 &  \\
Q1440$+$3539   &  0.077 &  14 40 04.6 &  $+$35 39 07  & 1440$+$3539              & 0.004 & 0.077 &  \\
Q1444$+$4047   &  0.267 &  14 44 50.2 &  $+$40 47 38  & PG1444$+$407             & 0.069 & 0.267 &  \\
Q1512$+$3701   &  0.371 &  15 12 47.4 &  $+$37 01 55  & B21512$+$37              & 0.330 & 0.371 &  W \\
Q1517$+$2356   &  1.903 &  15 17 08.3 &  $+$23 56 53  & LB9612                   & 1.449 & 1.694 &  \\
Q1517$+$2357   &  1.834 &  15 17 02.2 &  $+$23 57 44  & LB9605                   & 1.391 & 1.694 &  \\
Q1521$+$1009   &  1.324 &  15 21 59.9 &  $+$10 09 02  & PG1522$+$101             & 0.962 & 1.324 &  \\
Q1538$+$4745   &  0.770 &  15 38 01.0 &  $+$47 45 10  & PG1538$+$477             & 0.493 & 0.770 &  W \\
Q1544$+$4855   &  0.400 &  15 44 00.2 &  $+$48 55 25  & 1543$+$4855              & 0.182 & 0.400 &  \\
Q1618$+$1743   &  0.555 &  16 18 07.4 &  $+$17 43 30  & 3C334.0                  & 0.315 & 0.555 &  W \\
Q1622$+$2352   &  0.927 &  16 22 32.3 &  $+$23 52 01  & 3C336.0                  & 0.627 & 0.927 &  \\
Q1626$+$5529   &  0.133 &  16 26 51.5 &  $+$55 29 04  & PG1626$+$554             & 0.002 & 0.133 &  \\
Q1630$+$3744   &  1.400 &  16 30 15.2 &  $+$37 44 08  & 1630$+$377               & 1.094 & 1.478 &  \\
Q1634$+$7037   &  1.337 &  16 34 51.8 &  $+$70 37 37  & PG1634$+$706             & 0.978 & 1.337 &  UW \\
Q1641$+$3954   &  0.595 &  16 41 17.7 &  $+$39 54 10  & 3C345                    & 0.346 & 0.595 &  \\
Q1704$+$6048   &  0.371 &  17 04 03.5 &  $+$60 48 30  & 3C351.0                  & 0.160 & 0.371 &  UW \\
Q1715$+$5331   &  1.940 &  17 15 30.3 &  $+$53 31 27  & PG1715$+$535             & 1.481 & 1.694 &  \\
Q1718$+$4807   &  1.084 &  17 18 17.9 &  $+$48 07 10  & PG1718$+$481             & 0.758 & 1.084 &  \\
Q1803$+$7827   &  0.680 &  18 03 39.4 &  $+$78 27 54  & S51803$+$78              & 0.417 & 0.680 &  \\
Q1821$+$6419   &  0.297 &  18 21 44.1 &  $+$64 19 32  & H1821$+$643              & 0.094 & 0.297 &  UW \\
Q1845$+$7943   &  0.056 &  18 45 37.5 &  $+$79 43 06  & 3C390.3                  & 0.002 & 0.056 &  \\
Q2112$+$0556   &  0.457 &  21 12 47.6 &  $+$05 56 10  & PG2112$+$059             & 0.309 & 0.457 &  \\
Q2128$-$1220   &  0.501 &  21 28 52.9 &  $-$12 20 21  & PKS2128$-$12             & 0.310 & 0.501 &  W \\
Q2135$-$1446   &  0.200 &  21 35 01.2 &  $-$14 46 27  & PKS2135$-$147            & 0.013 & 0.200 &  \\
Q2141$+$1730   &  0.213 &  21 41 13.9 &  $+$17 30 02  & OX1692141$+$175          & 0.023 & 0.213 &  \\
Q2145$+$0643   &  0.990 &  21 45 36.2 &  $+$06 43 41  & PKS2145$+$067            & 0.679 & 0.990 &  UW \\
Q2243$-$1222   &  0.630 &  22 43 39.9 &  $-$12 22 40  & PKS2243$-$123            & 0.376 & 0.630 &  W \\
Q2251$-$1750   &  0.068 &  22 51 25.9 &  $-$17 50 54  & MR2251$-$178             & 0.002 & 0.068 &  \\
Q2251$+$1120   &  0.323 &  22 51 40.7 &  $+$11 20 40  & PKS2251$+$113            & 0.116 & 0.323 &  W \\
Q2251$+$1552   &  0.859 &  22 51 29.7 &  $+$15 52 54  & 3C454.3                  & 0.569 & 0.859 &  UW \\
Q2300$-$6823   &  0.512 &  23 00 27.9 &  $-$68 23 47  & PKS2300$-$683            & 0.309 & 0.512 &  W \\
Q2340$-$0339   &  0.896 &  23 40 22.6 &  $-$03 39 05  & PKS2340$-$036            & 0.600 & 0.896 &  W \\
Q2344$+$0914   &  0.672 &  23 44 03.9 &  $+$09 14 06  & PKS2344$+$092            & 0.411 & 0.672 &  W \\
Q2352$-$3414   &  0.706 &  23 52 50.8 &  $-$34 14 37  & PKS2352$-$342            & 0.439 & 0.706 &  W \\
\enddata

\tablenotetext{a}{Lower end of the analyzed redshift range.}
\tablenotetext{b}{Upper end of the analyzed redshift range.}
\tablenotetext{c}{U: object in Ulmer 1996 sample. V: object in Vanden
Berk et al.\ 1999 sample. W: object in Weymann et al.\ 1998 sample.}
\end{deluxetable}


\clearpage

\begin{deluxetable}{cccccccccc}
\tabletypesize{\scriptsize}
\tablewidth{0pt}
\tablecaption{Maximum likelihood fits for the HST/FOS sample.\label{tab:dndz-hst}}
\tablehead{%
\colhead{A/F\tablenotemark{a}} &
\colhead{$W_{\rm thr}$\tablenotemark{b}} &
\colhead{$W_{\rm max}$\tablenotemark{c}} &
\colhead{$V_{\rm ej}$\tablenotemark{d}} &
\colhead{$\Delta z$\tablenotemark{e}} &
\colhead{${\cal N}$} &
\colhead{$\gamma$}&
\colhead{$W_\star$} &
\colhead{${\cal A}_0$\tablenotemark{f}}  &
\colhead{$P_{\rm KS}$\tablenotemark{g}} \\
\colhead{\strut} &
\colhead{[\AA]} &
\colhead{[\AA]} &
\colhead{[km\thinspace s$^{-1}$]} &
\colhead{\strut} &
\colhead{\strut} &
\colhead{\strut} &
\colhead{[\AA]} &
\colhead{\strut} &
\colhead{\strut}
}
\startdata
A &  VAR & $\infty$ &    0 & 32.96 & 1298 & 0.65$\pm$0.12 & 0.300$\pm$0.008 & \nodata & 0.85 \\
A &  VAR & $\infty$ & 3000 & 29.48 & 1147 & 0.73$\pm$0.13 & 0.293$\pm$0.008 & \nodata & 0.74 \\
F &  VAR & $\infty$ &    0 & 32.45 & 1157 & 0.62$\pm$0.13 & 0.229$\pm$0.006 & \nodata & 0.88 \\
F &  VAR & $\infty$ & 3000 & 29.07 & 1038 & 0.60$\pm$0.14 & 0.229$\pm$0.006 & \nodata & 0.88 \\
A & 0.24 & $\infty$ & 3000 & 14.24 &  622 & 0.42$\pm$0.20 & 0.309$\pm$0.012 &    32.7 & 0.96 \\
F & 0.24 & $\infty$ & 3000 & 13.93 &  544 & 0.54$\pm$0.21 & 0.220$\pm$0.009 &    26.9 & 0.99 \\
F & 0.24 &     0.36 & 3000 & 13.93 &  223 & 0.38$\pm$0.32 & \nodata         &    12.3 & 0.56 \\
F & 0.36 & $\infty$ & 3000 & 19.98 &  437 & 0.62$\pm$0.21 & 0.218$\pm$0.010 &    25.8 & 0.93 \\
\enddata

\tablenotetext{a}{A: all Ly-$\alpha$ lines (forest and metal systems);
F: Ly-$\alpha$ forest only.}
\tablenotetext{b}{Rest equivalent width threshold.}
\tablenotetext{c}{Upper limit for rest equivalent width.}
\tablenotetext{d}{Min.\ ejection velocity.}
\tablenotetext{e}{Combined redshift path length in the sample.}
\tablenotetext{f}{Scaled to refer to lines with $\Wrest \geq
0.24$~\AA.}
\tablenotetext{g}{Kolmogorov-Smirnoff probability for the fit.}

\end{deluxetable}


\end{document}